\documentclass[journal,twoside,web]{ieeecolor}
\usepackage{makecell} % add this in preamble
\usepackage{generic}
\usepackage{textcomp}
\def\BibTeX{{\rm B\kern-.05em{\sc i\kern-.025em b}\kern-.08em
    T\kern-.1667em\lower.7ex\hbox{E}\kern-.125emX}}
\UseRawInputEncoding

\usepackage{siunitx}

\usepackage{float}
\usepackage{cite}

\usepackage{subcaption}

\usepackage{amsmath}

\usepackage{algorithmic}

\usepackage{array}

\usepackage{fixltx2e}

 \usepackage{dblfloatfix}
\usepackage{url}
\usepackage{graphicx}
\usepackage{amsmath}
\usepackage{fixmath}
\usepackage[hidelinks]{hyperref}
\usepackage{cleveref}  %for \cref i.e. multiple reference of tables in single reference
\usepackage{multirow,booktabs}
\usepackage{array}
\usepackage{tabularx}

\renewcommand{\tabcolsep}{1pt}
\newcolumntype{C}[1]{>{\centering\let\newline\\\arraybackslash\hspace{0pt}}m{#1}}
\newcolumntype{L}[1]{>{\raggedright\let\newline\\\arraybackslash\hspace{0pt}}m{#1}}

\def\ensuremath#1{\mbox{$#1$}}

\newcommand{\Pves}{\ensuremath{P_{ves}}}
\newcommand{\Pabd}{\ensuremath{P_{abd}}}
\newcommand{\Pdet}{\ensuremath{P_{det}}}

\usepackage{lineno,color}
\begin{document}

\bstctlcite{IEEEexample:BSTcontrol}

%\linenumbers   % after \begin{document} for line numbers

\title{
Automated Detection of Urological Events in Bladder Pressure Signals with a Two-Stage Machine Learning Framework Validated on External Datasets
}

\author{Hassaan A. Bukhari$^{1}$, 
    Vikram Abbaraju$^{2}$, 
    Jay Patel$^{3,4}$, 
    Becky Clarkson$^{5}$, 
    Shachi Tyagi$^{5}$, 
    Margot S. Damaser$^{1,4,6}$,
    Steve J. A. Majerus$^{3,4}$ \\
\ \\ % leave an empty line between authors and affiliation
$^1$ Department of Biomedical Engineering, Cleveland Clinic Research, Cleveland, OH, USA\\
$^2$ School of Electrical and Computer Engineering, Georgia Institute of Technology, Atlanta, GA, USA\\
$^3$ Case Western Reserve University, Cleveland, OH, USA\\
$^4$ Advanced Platform Technology Center, Louis Stokes Cleveland Veterans Affairs Medical Center, Cleveland, OH, USA \\
$^5$ Division of Geriatric Medicine, University of Pittsburgh, Pittsburgh, PA, USA \\
$^6$ Glickman Urology Institute, Cleveland Clinic, Cleveland, OH, USA\\}

\markboth{Engineering Applications of Artificial Intelligence}%
%\markboth{manuscript in preparation}%
{Bukhari \MakeLowercase{\textit{et al.}}: Automated Detection of Urological Events}

\maketitle

%{\renewcommand{\thefootnote}{}\footnotetext{Copyright (c) 2026 IEEE. Personal use of this material is permitted. However, permission to use this material for any other purposes must be obtained from the IEEE by sending an email to pubs-permissions@ieee.org.}}

\begin{abstract} % max 250 words
%\vspace{-0.2cm}

\boldmath
\textbf{\textit{Objective:}} Conventional urodynamics (UDS) provide critical diagnostic information, but require invasive dual catheterization and manual labeling of clinically important events. Wireless, catheter-free bladder function tests are becoming available for home use, but only provide vesical pressure (\Pves). We developed a machine learning framework that was trained and externally validated on UDS data for automated urological event classification from single-channel (\Pves) recordings.
\textbf{\textit{Methods:}} We analyzed 118 annotated UDS traces segmented into 0.8-second \Pves\ intervals. Using the discrete wavelet transform, we extracted 55 statistical features per segment. Consecutive segments (233,338 segments; three classes) sharing the same class, abdominal (ABD), detrusor overactivity (DO), or voiding contraction (VOID), were grouped into events, and median feature aggregation was applied to derive event-level representations. Using an imbalanced dataset, we trained a two-stage multilayer perceptron (MLP): Stage 1 distinguished VOID vs non-VOID, and Stage 2 classified non-VOID into ABD and DO. The model was trained on two independent datasets and externally validated on a third independent dataset. Additional cross-dataset training-–validation permutations were performed to assess generalizability. Performance was evaluated using accuracy, F1-macro, sensitivity, specificity, and area under the curve (AUC).
\textbf{\textit{Results:}} Stage 1 (VOID vs. non-VOID) achieved 84\% accuracy (balanced accuracy 76\%), F1-macro 0.74, and AUC 0.85, while Stage 2 (ABD vs. DO) reached 90\% accuracy (balanced accuracy 80\%), F1-macro 0.80, and AUC 0.87. Permutation feature importance indicated that most features contributed meaningfully.
\textbf{\textit{Conclusion:}}  Our machine learning approach enables accurate automated detection of urological events from \Pves, demonstrating feasibility for single-channel monitoring and future ambulatory applications. 
%with potential to improve access to urological care.
  
\unboldmath
\end{abstract}


% \fi
\IEEEpeerreviewmaketitle

\section{Introduction}
\IEEEPARstart{L}{ower} urinary tract dysfunction (LUTD) represents a global health burden, with an estimated 45\% of the world population and 20\% of the US adult population being affected, significantly impacting their quality of life with symptoms such as increased urinary urgency, frequency, or incontinence \cite{maserejian_incidence_2013,abrams_standardisation_2002,irwin_worldwide_2011,zhang_prevalence_2018}.  Urodynamic studies (UDS) are the clinical standard for evaluating lower urinary tract function and dysfunction \cite{drake_fundamentals_2018,abrams_standardisation_2002}, and are used to assess conditions including neurogenic lower urinary tract dysfunction (NLUTD) and overactive bladder (OAB).

Conventional UDS uses two catheters, one to measure bladder pressure (\Pves) and another for abdominal pressure (\Pabd), in order to derive detrusor pressure (\Pdet = \Pves $-$ \Pabd), while also recording bladder volume, flow, and patient sensations to identify events such as abdominal activity (ABD), detrusor overactivity (DO), and voiding contractions (VOID) \cite{drake_fundamentals_2018,nitti_practical_1998}. However, interpretation is often subjective, requires training, and can be affected by artifacts, errors, catheter displacement, and lack of standardization \cite{pourghazi_artifacts_2025}. Furthermore, the UDS catheter instrumentation can cause patient discomfort and limit the ability to reproduce typical patient symptoms\cite{finkelstein_anxiety_2020,oktaviani_intraurethral_2021,vogt_catheter-free_2024}. 

These limitations have driven interest in ambulatory and catheter-free monitoring, particularly single-sensor approaches that measure \Pves\ only, and in automated computational methods that detect urological events from \Pves\ alone, enabling home-based monitoring, telemetric ambulatory urodynamics, and closed-loop neuromodulation \cite{karam_real-time_2016,majerus_feasibility_2021,abelson_ambulatory_2019}. While \Pves\ reflects both bladder and abdominal activity and can guide clinical decisions \cite{ginsberg_auasufu_2021}, current wireless, catheter-free ambulatory systems measure only \Pves\ \cite{frainey_first_2023,gross_validation_2026,kim_feasibility_2025}, making differentiation of critical events such as ABD, DO, and VOID challenging. Therefore, accurate, automated analysis of \Pves\ to reliably detect these events is of great clinical importance, supporting timely diagnosis, improved patient care, and effective use of home-based ambulatory technology.

Prior efforts in automated event detection from \Pves\ include the Context-Aware Thresholding (CAT) algorithm which achieved approximately 97\% accuracy in detecting bladder contractions and motion artifacts \cite{karam_real-time_2016, majerus_feasibility_2021}, but was limited to binary classification and could not differentiate contraction types or distinguish motion from abdominal events. While such threshold-based approaches are effective for detecting general activity, they cannot model the complex temporal and morphological patterns that characterize different urodynamic events. Machine learning (ML) methods, in contrast, can learn discriminative features directly from data, enabling more specific classification of urological events and supporting real-time annotation in ambulatory settings.

ML strategies have increasingly been applied to urodynamic data to detect specific event types. Early efforts such as Wang et al. \cite{wang_pattern_2021} used manifold learning to detect DO from \Pdet, achieving around 81\% accuracy but requiring multiple pressure inputs (\Pves\ and \Pabd) and addressing only a single event. Hobbs et al. \cite{hobbs_machine_2022} extracted features from \Pves\ and multi-channel recordings (\Pves, \Pabd, \Pdet), reaching up to 92.9\% specificity for DO with three channels (and 70.3\% specificity with only \Pves), yet relied on long 60-second windows, limiting real-time use. More recent studies, including Mei et al. \cite{mei_deep_2025} and Liu et al. \cite{liu_real-time_2025}, demonstrated high diagnostic accuracy for urological events using deep learning, but both relied on multi-channel recordings, limiting applicability for single-sensor or ambulatory settings. Recently published narrative and systematic reviews have highlighted the promise of ML in urodynamics, but most studies remain preliminary, often single-center, focused on only DO detection, and are not yet suitable for routine clinical use \cite{liu_applications_2024,gammie_can_2024,chew_future_2025}. To overcome these limitations, an autonomous bladder pressure event classifier (ABEC) was developed in our group as a real-time, single-channel \Pves\ classifier for ABD, DO, and VOID \cite{abbaraju_real-time_2023,majerus_real-time_2024}, although challenges remain in misclassification, noisy segments, and segment-independent predictions.

To address these shortcomings, we propose a two-stage neural network and further improve the ABEC framework to detect urological events from single-channel \Pves. The first stage discriminates VOID versus non-VOID events, while the second stage performs fine-grained classification among ABD and DO. This hierarchical approach integrates thresholding and ML to improve accuracy, specificity, and sensitivity, reduce misclassification, and enhance real-time clinical applicability for \Pves\ analysis. The model was developed and externally validated using UDS \Pves\ recordings from three independent datasets, with proof-of-concept evaluation performed on ambulatory \Pves\ recordings.

\section{Materials and Methods}
\label{Methods}
\subsection{Study population and data preparation}
\begin{figure*}[h]% a [!h] places the figure in the middle of a colum and
              % takes twice the vertical separation space
  \centering
    \includegraphics[width=0.95\textwidth]{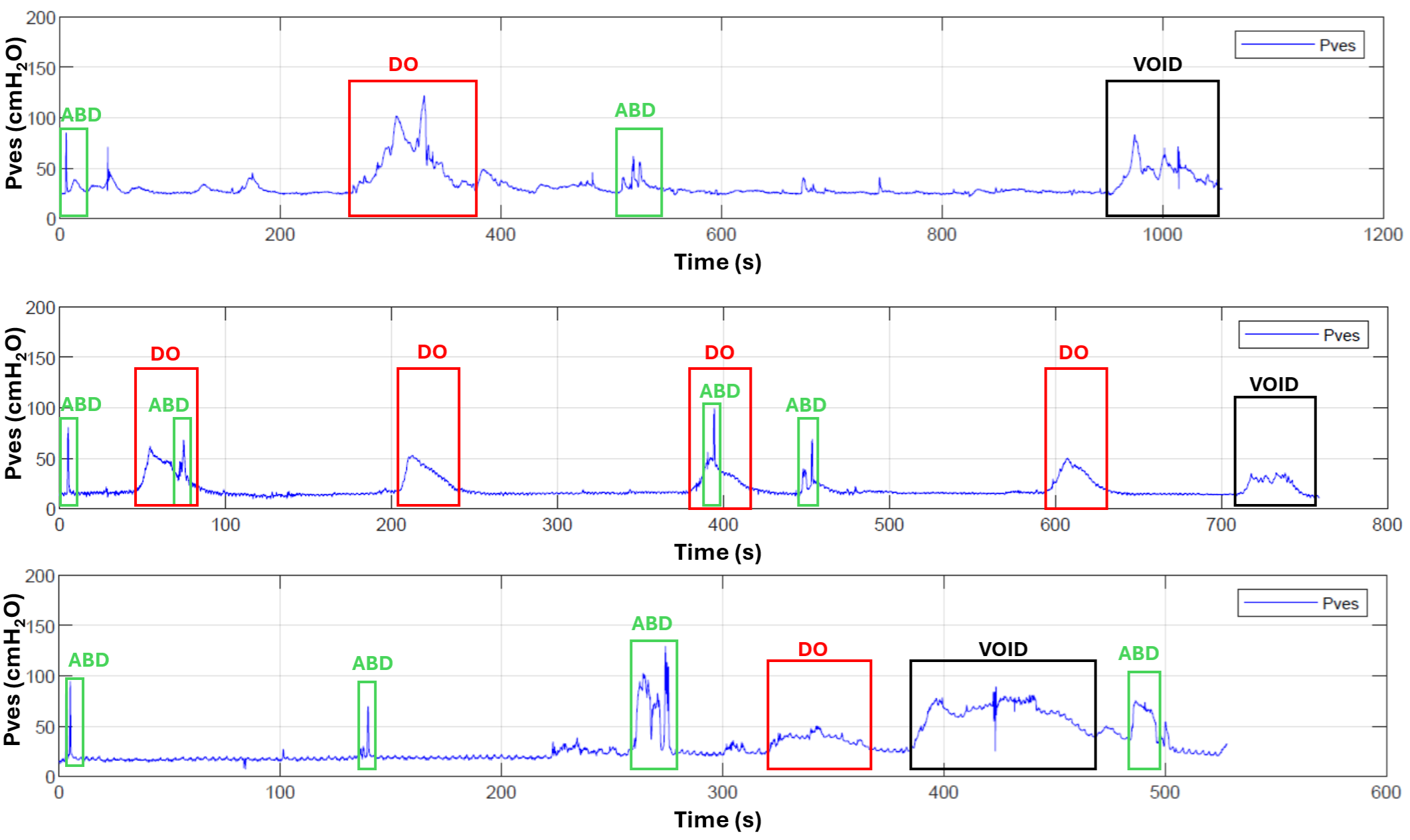}
   % \put(-104,41){\footnotesize\textcolor{red}{/215}}
  \caption{Three representative \Pves\ (vesical pressure) signals manually annotated for voiding contractions (Void, black), abdominal activity (ABD, green), and detrusor overactivity (DO, red). These event annotations were used as ground truth for training and evaluating the machine learning models.}
 \label{Pves_annotated_urologist} 
\end{figure*}

The study population included 118 pre-recorded UDS traces from 76 human patients with overactive bladder or neurogenic urinary incontinence, collected from the Louis Stokes Cleveland Department of Veteran Affairs Medical Center and the Cleveland Clinic. These included 64 traces from female patients (Dataset A) with neurogenic bladder, 20 traces from female patients (Dataset B) with overactive bladder, and 34 traces from male patients  (Dataset C) with neurogenic bladder following spinal cord injury (SCI). Of these, 98 traces were originally sampled at 100 Hz, while the remaining 20 traces were sampled at 10 Hz. To ensure consistency in temporal resolution across datasets, all recordings were downsampled to a uniform sampling rate of 10 Hz prior to analysis. Each UDS trace included \Pves, \Pabd, \Pdet, bladder volume, and flow rate. An expert urologist reviewed the entire UDS trace to annotate three urological events: ABD, DO, and VOID, while the remaining portions were considered as No-event (NONE). For this study, only the \Pves\ data along with the annotated labels were used for analysis (Fig. \ref{Pves_annotated_urologist}).  We used pre-existing, fully de-identified data from earlier studies that did not require formal IRB approval.

%Table \ref{tab:HDECG} shows the population characteristics.   

%\begin{table}[h]
%\caption{\label{tab:HDECG} Characteristics of the 108 UDS study population. Values are expressed as number ($\%$) for categorical variables and median (interquartile range) for continuous variables.} 

%%\small\addtolength{\tabcolsep}{3pt}
%\centerline{\begin{tabular}{L{5cm}r} \hline\hline
%Characteristics    & Quantity  \\ \hline
% \ \ Age [years]  &    $72\ (10)$ \\
% \ \ Gender [male/female]  &    14/6   \vspace{0.15cm}   \\
% %\hline
 
% {\bf Electrolyte concentrations }    & \\
%%median (IQR)
%  \ \ \kaco{} [Pre HD] (mM) &   $5.36\ (1.69)$ \\
%  \ \ \kaco{} [End HD] (mM) &   $3.25\ (0.63)$   \\
%  \ \ \Caco{} [Pre HD] (mM) &   $2.15\ (0.17)$  \\
% \ \  \Caco{} [End HD] (mM) &   $2.35\ (0.17)$   \\ \hline
%\vspace{5pt}
%     & \#Patients  ($\%$)\\ \hline
%{\bf HD session duration}     & \\ 
% \ \  240 min & $17\ (85\%)$ \\
% \ \  210 min & $3\ (15\%)$ \\
%{\bf Dialysate composition} & \\
%  \ \ Potassium (1.5 mEq/L) &  $17\ (85\%)$  \\
%  \ \ Potassium (variable mEq/L) &  $3\ (15\%)$  \\
%  \ \ Calcium (1.5 mEq/L) &  $6\ (30\%)$    \\
%  \ \ Calcium (1.25 mEq/L) &  $14\ (70\%)$ \\

%\hline\hline
%\end{tabular}}
%\end{table}

%\vspace{5pt}
%\section{Methods}

\par
\begin{figure}[htbp]
    \centering
    \includegraphics[width=0.49\textwidth]{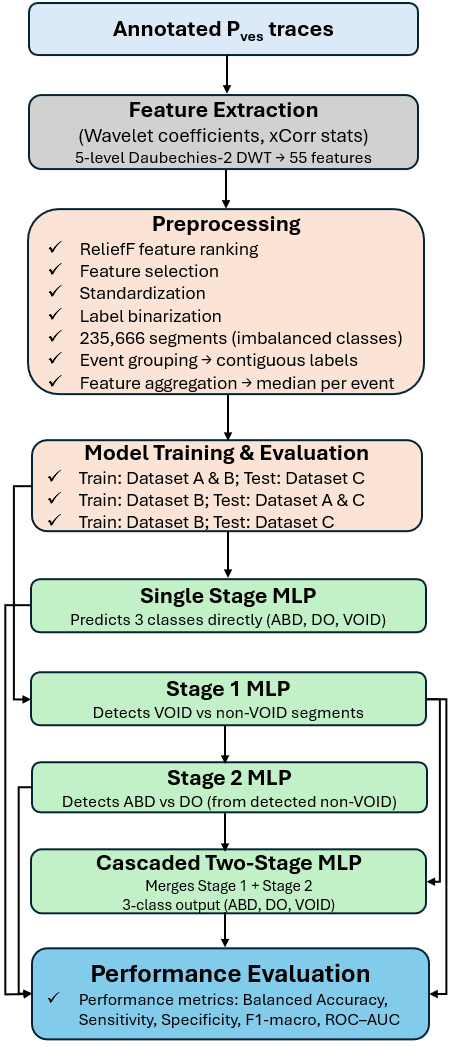}
    \caption{Overview of the data processing and classification workflow, including manual \Pves\ annotation, preprocessing, feature extraction, single and two-stage classification, and evaluation metrics computation. Abbreviations: \Pves, vesical pressure; xCorr, cross-correlation; DWT, discrete wavelet transform; MLP, multilayer perceptron; ABD, abdominal activity; DO, detrusor overactivity; VOID, voiding contraction; ROC–-AUC, area under the receiver operating characteristic curve.}
    \label{fig:workflow}
\end{figure}

An overview of the complete data processing and classification workflow is shown in Fig.~\ref{fig:workflow}. The pipeline starts with manual annotation of \Pves\ traces, followed by preprocessing and feature extraction. A two-stage classification is then performed, and results are evaluated using accuracy, confusion matrices, and receiver operating characteristic (ROC) analysis.

\subsection{Feature extraction}
\label{subsection}

ABD, DO and VOID events show distinct time-frequency signatures: ABD events are short, high amplitude spikes (transient pressure spikes), while DO and VOID events are low-frequency gradual rises and falls in pressure. More specifically, DO events are characterized by gradual rises and falls in pressure and VOID events represent sustained increases in pressure followed by a return to baseline \cite{drake_fundamentals_2018,nitti_practical_1998}. To detect these patterns via machine learning, we thus extracted frequency-domain features. Traditional approaches, such as the fast Fourier transform (FFT) or short-time Fourier transform (STFT), have limitations for non-stationary signals like \Pves\ due to poor temporal localization or trade-offs between time and frequency resolution \cite{dirgenali_estimation_2006,zhang_comparison_2003}. In this study, the discrete wavelet transform (DWT) was used for multi-resolution analysis, offering improved time-–frequency localization and computational efficiency \cite{nazeran_wavelet-based_2007,karam_real-time_2016,batista_miranda_2025}. Features computed using the DWT applied on \Pves\ data formed the basis for training the machine learning classifiers.

\subsubsection{Time-frequency analysis using DWT} 
We applied a 5-level DWT using the Daubechies-2 (Db2) mother wavelet to extract meaningful features from the non-stationary \Pves\ signals \cite{karam_real-time_2016}. The Db2 wavelet–-based DWT decomposes the signal into approximation and detail coefficients  at each level through low-pass and high-pass filtering followed by downsampling. This multi-resolution decomposition allows the DWT to capture both low-frequency trends and the precise timing of high-frequency events. A cubic spline interpolation was applied to each coefficient set to match the original signal length to preserve temporal resolution across all DWT levels \cite{abbaraju_real-time_2023}.

\subsubsection{Statistical features from DWT coefficients} 
After applying DWT and interpolation, the signal was segmented into 0.8-second non-overlapping windows (8 \Pves\ samples per segment at 10 Hz). This window length, slightly shorter than 1 second, was selected as a power of 2 to simplify downstream computation while maintaining adequate temporal resolution. Each \Pves\ segment was labeled with each of the class labels, ABD, DO, VOID, and NONE, depending on which event was present in each segment. From each of the 10 wavelet coefficient signals obtained from the five-level DWT decomposition (five approximation coefficients, cA1-–cA5, and five detail coefficients, cD1–-cD5), four statistical features \cite{chashmi_efficient_2019, cvetkovic_wavelet_2008}, maximum, mean absolute value, median, and Shannon entropy, were computed, resulting in 40 wavelet-based features (e.g., $cA5_{max}$, $cA5_{mav}$, $cA5_{med}$, $cA5_{ent}$). In addition, the cross-correlation between approximation and detail coefficients at each DWT level (i = 1--5) was computed, and the maximum, mean, and median of each cross-correlation signal were extracted, resulting in 15 additional features (e.g., $xCorr3_{max}$, $xCorr3_{mean}$, $xCorr3_{med}$). In total, 55 features were extracted from each segment, resulting in 233,338 \Pves\ segments across the dataset and forming an input matrix of size $233{,}338 \times 55$. The dataset was highly imbalanced, with the majority of segments labeled as NONE. To mitigate this, 120,161 segments were randomly selected for the NONE class, approximately matching the combined total of ABD (23,835), DO (22,748), and VOID (71,250) segments. These segment-level features were subsequently used to derive event-level representations for training and evaluating the neural network classifier.

\subsubsection{Event-level feature aggregation and classification}
Consecutive segments sharing the same class label (ABD, DO, VOID) were grouped into events. For each event, features were aggregated using the median across all constituent segments to reduce noise and summarize event-level characteristics. Event-level feature vectors were then used to train a neural network classifier. For visualization, predicted event labels were propagated back to all constituent segments along the original Pves waveform. Segments labeled as ``NONE" (representing baseline periods without a physiological event) were excluded from the training and evaluation of the event classification model so that the analysis focused on clinically relevant events (ABD, DO, and VOID). Table~\ref{tab:event_counts} shows the number of distinct ABD, DO, and VOID events in each dataset after grouping consecutive segments of the same class into single events.

\begin{table}[h!]
\centering
\caption{Number of ABD, DO, and VOID events per dataset after aggregation.}
\renewcommand{\arraystretch}{1.5} % keeps row height same
\small
\begin{tabular}{lcccccc}
\hline
\textbf{Dataset} && \textbf{ABD} && \textbf{DO} && \textbf{VOID} \\
\hline
A && 78 && 23 && 59 \\
B && 118 && 29 && 17 \\
C && 150 && 40 && 37 \\
\hline
\end{tabular}
\label{tab:event_counts}
\end{table}

\subsection{Artificial Neural Network Model, Training, and External Validation}
A fully connected multilayer perceptron (MLP), which is a type of artificial neural network (ANN), was implemented to classify \Pves\ segments into ABD, DO, or VOID events. The MLP architecture was selected to capture nonlinear relationships in the extracted feature space while maintaining relatively low model complexity and reducing the risk of overfitting \cite{rossi_functional_2005,khashei_novel_2012,przybyla-kasperek_multi-layer_2024}. The network comprised an input layer of 55 neurons corresponding to the extracted 55 statistical and wavelet-based features, two hidden layers with 128 and 200 neurons with rectified linear unit (ReLU) activations, and an output layer with two neurons for binary classification in each stage (Stage 1: VOID vs non-VOID, Stage 2: ABD vs DO) followed by a softmax function to generate class probabilities. Feature selection using the ReliefF algorithm, a widely used feature ranking method for estimating feature relevance, was applied to identify features that best distinguish between classes \cite{urbanowicz_relief-based_2018,kononenko_estimating_1994}. Input features were standardized using z-score normalization by subtracting each feature by mean and dividing by its standard deviation \cite{bishop_pattern_2006}. The scaler was fit on the training data and subsequently applied to the test set to ensure consistent scaling and avoid data leakage. Hyperparameter selection was performed using k-fold cross-validation on the training data to optimize model performance.

The MLP was trained using the Adam optimizer with a learning rate of 0.001 and a cross-entropy loss function. Training was performed for 50 epochs with a batch size of 128. Model training and evaluation were conducted on a machine equipped with an Intel CPU (12 cores, 20 logical processors, base speed 2.10 GHz), 16 GB of RAM, and a GPU with 15.8 GB of memory. Training and evaluation of the MLP took less than one minute per experiment.

To evaluate model performance and generalizability, three complementary validation strategies were applied. First, internal validation used a stratified random split of the pooled datasets (A, B, and C), with 60\% for training and 40\% for testing. Second, external validation trained the model on independent datasets A and B and tested on independent dataset C, which remained entirely unseen, assessing robustness to distribution shift. Third, to examine performance under limited training conditions and heterogeneous testing, the model was trained solely on dataset B and evaluated on the combined datasets A and C.

Additional training-–testing configurations, including 80\%–-20\% and 70\%-–30\% random splits, as well as training on dataset B with testing only on dataset C, were also explored. Results for these supplementary experiments are provided in the supplementary material. To further prevent data leakage, even in random splits, all events from the same \Pves\ trace were assigned exclusively to either the training or testing set.

\paragraph{Two-Stage Hierarchical Classification}
A two-stage hierarchical model was designed. The first stage classifies between `VOID' and `non-VOID'. The second stage operates only on samples identified as `non-VOID' and assigns them to either ABD, or DO. 
%The predicted class corresponds to the highest softmax probability, enabling the network to learn nonlinear decision boundaries that effectively separate all classes.

\subsubsection{Comparison Configurations}
For comparison, two additional MLP configurations were also evaluated:

\begin{enumerate}
    \item \textit{Cascaded two-stage MLP}: Stage~1 (VOID vs.\ non-VOID detection) and Stage~2 (ABD and DO classification) were combined to produce a unified three-class output (ABD, DO, VOID). 
    \item \textit{Single-stage MLP}: directly classified three classes (ABD, DO, VOID) in a single step.
\end{enumerate}

Both configurations used the same network architecture, training strategy, and hyperparameter optimization procedure as the main two-stage model, and their performance was compared using identical evaluation metrics. 
%These comparisons were conducted to assess whether the proposed hierarchical approach offers improved accuracy and generalization compared to conventional single-stage architectures.

\subsubsection{Evaluation Metrics}
Model performance was evaluated using standard metrics such as balanced accuracy, sensitivity (recall), specificity, overall accuracy, and F1-macro \cite{powers_evaluation_2020}. ROC curves and AUC were computed per class, and confusion matrices were visualized using heatmaps \cite{fawcett_introduction_2006}. 

%The metrics are defined as follows :

%\begin{align}
%    \text{Accuracy} &= \frac{TP + TN}{TP + TN + FP + FN} \\
%    \text{Sensitivity (Recall)} &= \frac{TP}{TP + FN} \\
%    \text{Specificity} &= \frac{TN}{TN + FP} \\
%    \text{Precision} &= \frac{TP}{TP + FP} \\
%    \text{F1-score} &= 2 \times \frac{\text{Precision} \times \text{Recall}}{\text{Precision} + \text{Recall}}
%\end{align}

%where \(TP\), \(TN\), \(FP\), and \(FN\) denote true positives, true negatives, false positives, and false negatives, respectively.

%The ROC curve plots the true positive rate (sensitivity) against the false positive rate (\(1 - \) specificity) at various threshold settings. The AUC summarizes the ROC curve as the probability that a classifier ranks a randomly chosen positive instance higher than a randomly chosen negative one \cite{fawcett_introduction_2006}. 

To assess the contribution of each feature to the MLP model's predictive performance, we employed the permutation feature importance (PFI) method \cite{maragno_predictive_2025,thomas_machine_2024,chen_investigating_2023,mi_permutation-based_2021}. PFI quantifies the increase in prediction error that occurs when a feature's values are randomly permuted, thereby disrupting its relationship with the target variable. Feature importance was measured as the corresponding drop in classification accuracy on the test set. A significant increase in error indicates that the model relies heavily on that feature for accurate predictions. PFI highlighted the feature with the most significant impact on the model.

\subsection{Proof-of-Concept Evaluation on UM Ambulatory Data}
To further demonstrate the feasibility of our algorithm on independent ambulatory data, in addition to evaluating performance on external datasets, the trained model from Datasets A and B was applied to a proof-of-concept evaluation using ambulatory \Pves\ data collected with our proprietary wireless catheter-–free bladder pressure measuring urodynamics monitor (UM) system \cite{frainey_first_2023,gross_validation_2026}. The objective of this proof-of-concept evaluation was to determine whether the models, trained exclusively on UDS recordings, could still recognize physiologically meaningful events when exposed to ambulatory recordings acquired from a different device and in a different environment. The study was approved by the Institutional Review Board (IRB) at the University of Pittsburgh for this research use (IRB number: STUDY22050099).

Continuous \Pves\ was recorded for approximately 24 hours from patients enrolled in an ongoing collaborative study between the Continence Research Center of the University of Pittsburgh, the Cleveland Clinic, and the Advanced Platform Technology (APT) Center at the Louis Stokes Cleveland VA Medical Center (LSVAMC). For this proof-of-concept, we included data from a single patient, a 61-year-old woman with 4-year history of voiding symptoms including frequency, urgency, urge urinary incontinence (urinary leakage with sudden onset of urgency), and nocturia (getting up at night to void) frequency of 2 each night. She was not taking any prescription or over-the-counter bladder medications (e.g., anticholinergics, antimuscarinics, or $\beta_3$ agonist) for at least 6 weeks prior to the overnight study procedure. The analyzed \Pves\ data included the sleep period itself, as well as the 60 minutes preceding sleep onset and the 60 minutes following waking.

For this preliminary analysis, the trained Stage 1 and Stage 2 MLP models were applied directly to the raw \Pves\ signals without preprocessing (e.g., filtering or segmentation). Although formal ground-truth event labels were not fully available for the UM dataset, selected segments with events were cross-checked against patient-reported bathroom activity to provide preliminary confirmation of physiologically meaningful events.

%, followed by two fully connected hidden layers, each containing 16 neurons and employing the Rectified Linear Unit (ReLU) activation function. The output layer consisted of four neurons with softmax activation to classify each segment into one of the four event categories: DO (detrusor overactivity), ABD (abdominal event), VOID, or NONE (no event).

%The model was implemented using the Keras API with TensorFlow backend. The network was compiled using the Adam optimizer and trained with the categorical cross-entropy loss function. Training was performed for 150 epochs with a batch size of 64. The model was trained using an 80/20 split of the balanced dataset (233,338 samples total; 4,000 per class), and performance was evaluated on the held-out test set using accuracy, precision, recall, and F1-score metrics.

%To mitigate overfitting, early stopping and dropout (if implemented) can also be described here if applicable. Additionally, model training and evaluation were conducted on a machine equipped with [insert hardware if needed, e.g., "NVIDIA GPU or Intel i7 CPU with 16 GB RAM"].

\section{Results}

%We assessed the proposed hierarchical framework consisting of a binary VOID vs. non-VOID detector (Stage~1), followed by a secondary classifier (Stage~2) that discriminates ABD from DO exclusively among samples predicted as non-VOID. To benchmark the hierarchical design, we additionally evaluated a single-stage three-class MLP and a unified cascaded configuration under identical training and evaluation protocols.

\subsection{Single Stage: ABD, DO, VOID Classification}

The single-stage classifier, which directly distinguished ABD, DO, and VOID events, demonstrated strong discriminative performance (AUC up to 0.87) across all evaluation scenarios. Performance was generally strong for ABD and VOID events (balanced accuracy up to 81), while classification of DO events was slightly more challenging (balanced accuracy up to 68), particularly in cross-dataset validation. Notably, when externally validated on dataset C, the model achieved an overall accuracy of 74\% and an AUC of 0.83. Detailed metrics for overall accuracy, F1-macro, AUC, and class-wise balanced accuracy, sensitivity, and specificity are provided in Tables~\ref{tab:stage1_stage2_metrics_40}, \ref{tab:stage1_stage2_metrics_ABC}, and \ref{tab:stage1_stage2_metrics_BAC}.

%\begin{figure}[ht]
%\centering
%\includegraphics[width=0.49\textwidth]{Figures_new/confusion_matrix_4class.png}
%\includegraphics[width=0.49\textwidth]{Figures_new/stage1_confusion_matrix_percent_relabel_data_v3.png}
%\caption{Confusion matrix (raw counts in left panel and normalized, in \%, in right panel) for Single Stage classification, on the unseen test set.}
%\label{fig:single_stage_cm_cm_percent_unseen_test_data}
%\end{figure}

\subsection{Stage 1: VOID vs non-VOID Classification}

In Stage 1 (VOID vs non-VOID), the model achieved an overall accuracy of \textbf{84\%} when validated on dataset C (\textbf{81\%} with 40\% unseen test set and \textbf{74\%} when validated on datasets A and C) (Tables~\ref{tab:stage1_stage2_metrics_40}, \ref{tab:stage1_stage2_metrics_ABC}, and \ref{tab:stage1_stage2_metrics_BAC}). The model in Stage~1 demonstrated good discriminative ability across all evaluation scenarios (VOID, non-VOID: 65\%, 88\% when validated on Dataset C, 90\%, 79\% with 40\% test, 88\%, 70\% when validated on Datasets A and C, respectively), with AUC values reaching 89\% (Fig.~\ref{fig:all_stages_cm_raw_unseen_test_data}, Figure~S1 in supplementary material).
%For comparison, we also provide the confusion matrix obtained from five-fold cross-validation on the training data in the supplementary material (Fig.~S1). 
%Fig.~\ref{fig:roc_stage1_stage2} (panel~b) shows the Stage 1 ROC curve for binary classification between NONE and EVENT classes.

%\begin{table}[ht]
%\centering
%\caption{Performance metrics for Stage 1 classification (NONE vs EVENT).}
%\label{tab:stage1_metrics}
%\begin{tabular} {|p{3cm}|p{2cm}|}
%\hline
%\textbf{Metric} & \textbf{Value} \\
%\hline
%Accuracy (\%) & 90.91 \\
%Sensitivity (\%) & 79.97 \\
%Specificity (\%) & 94.59 \\
%Precision & 0.83 \\
%F1 Score & 0.82 \\
%Mean AUC (EVENT) & 0.94 \\
%AUC StdDev (EVENT) & 0.002 \\
%\hline
%\end{tabular}
%\end{table}

%\begin{figure}[ht]
%\centering
%\includegraphics[width=0.49\textwidth]{Figures_new/stage1_confusion_matrix_counts_percent_v9_imbalance.png}
%%\includegraphics[width=0.49\textwidth]{Figures_new/stage1_confusion_matrix_percent_relabel_data_v3.png}
%\caption{Confusion matrix (raw counts in left panel and normalized, in \%, in right panel) for Stage 1 classification, on the unseen test set.}
%\label{fig:stage1_cm_cm_percent_unseen_test_data}
%\end{figure}

%Fig. \ref{fig:PFI_stage1_stage2} (left panel) illustrates the feature importance rankings for Stage 1 model.  The features are ordered by their impact on model accuracy, with higher values indicating greater importance. 
%The ANN assigned the highest PFI to the cross-correlation between approximation and detail wavelet coefficients, indicating their critical contribution to the model’s predictive capability.

For Stage 1, permutation feature importance analysis indicated that most computed features contributed meaningfully to classification performance, with the top 10 features exhibiting substantial influence on model predictions (Fig.~\ref{fig:PFI_stage1_stage2}, and Figure~S2, panel~a).

\subsection{Stage 2: ABD vs DO Classification}

\begin{figure*}[ht]
\centering
\includegraphics[width=0.88\textwidth]{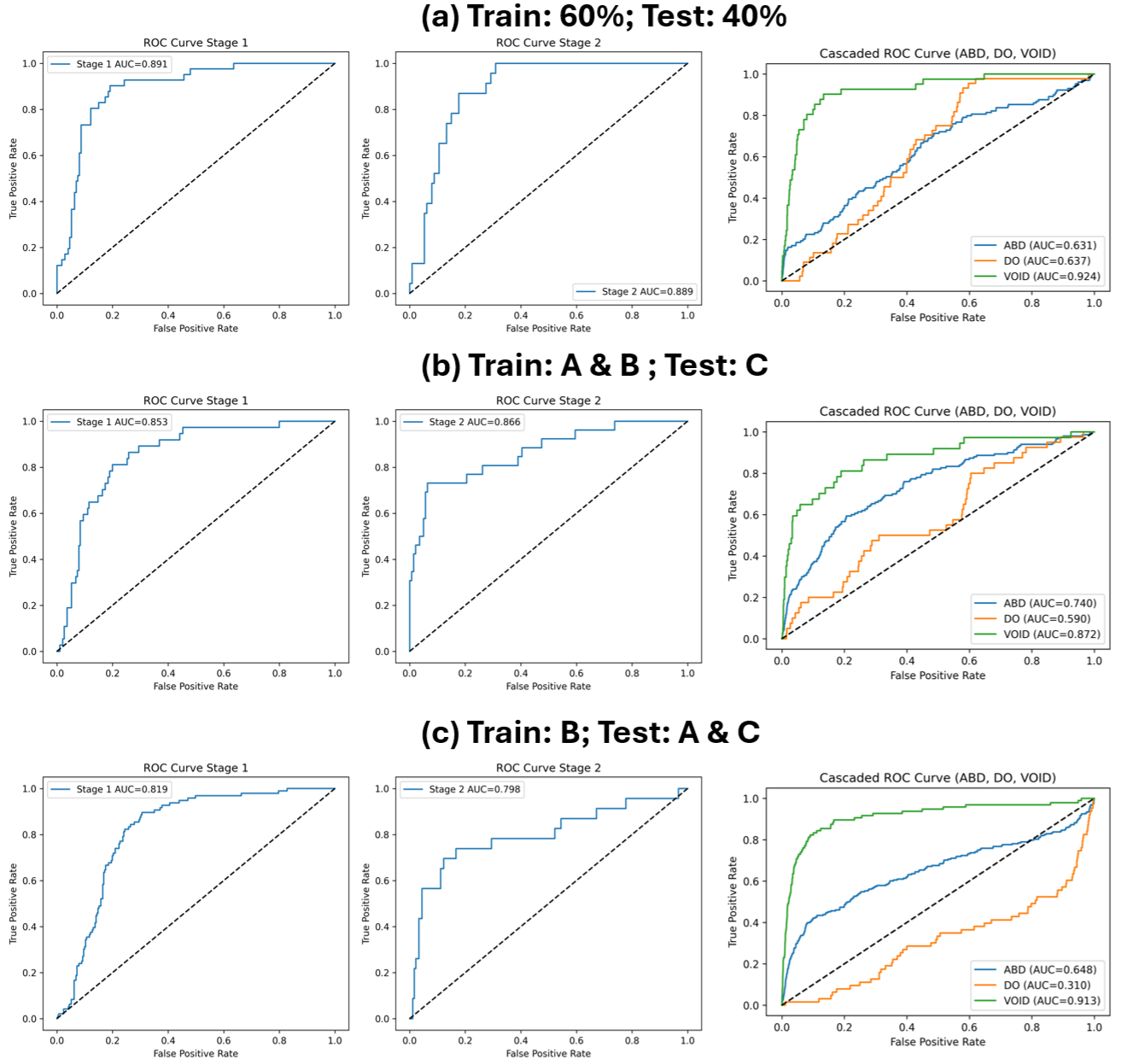}
\caption{ROC curves for the full classification pipeline. Panels (a–c) show performance under different training and testing configurations: (a) training on 60\% and testing on 40\% of the data, (b) training on Datasets A and B and testing on Dataset C, and (c) training on Dataset B and testing on Datasets A and C. Within each panel, ROC curves are shown for Stage 1 (left), Stage 2 (middle), and the cascaded Stage 1+2 system (right), with each plot displaying the mean ROC curve and the corresponding mean AUC values.}
\label{fig:all_stages_cm_raw_unseen_test_data}
\end{figure*}

In Stage 2, samples identified as non-VOID were further classified into ABD and DO. We achieved an overall accuracy of \textbf{90\%} when validating on dataset C, \textbf{82\%} when validating on the 40\% unseen test set, and \textbf{81\%} when validating on datasets A and C (Tables~\ref{tab:stage1_stage2_metrics_40}, \ref{tab:stage1_stage2_metrics_ABC}, and \ref{tab:stage1_stage2_metrics_BAC}). We achieved strong per-class discrimination, with accuracies of 94\% for ABD, and 65\% for DO when validated on dataset C (ABD, DO: 82\%, 83\% with 40\% unseen test set and 82\%, 74\% when validated on datasets A and C)  (Figure~S1 in supplementary material). ROC analysis further confirmed good separability among the two event classes, with AUC values reaching 89\% across two classes (Fig.~\ref{fig:all_stages_cm_raw_unseen_test_data}).

%40% unseen dataset

\begin{table}[ht]
\centering
\caption{Performance metrics for Single Stage (ABD, DO, VOID), Stage 1 (VOID vs Non-VOID), 
Stage 2 (ABD vs DO), and Cascaded Stage 1+2 (ABD, DO, VOID) 
using the MLP classifier on the 40\% unseen test set (Train:60\%, Test:40\%).}
\label{tab:stage1_stage2_metrics_40}
\renewcommand{\arraystretch}{1.3}
\setlength{\extrarowheight}{1pt}
\resizebox{0.49\textwidth}{!}{%
\begin{tabular}{|l|c|c|c|c|c|c|c|c|c|c|}
\hline
\multirow{2}{*}{\textbf{Metric(Tr:60\%,Te:40\%)}} 
& \multicolumn{3}{c|}{\textbf{Single Stage}} 
& \multicolumn{2}{c|}{\textbf{Stage 1}} 
& \multicolumn{2}{c|}{\textbf{Stage 2}} 
& \multicolumn{3}{c|}{\textbf{Cascaded Stage}} \\
\cline{2-11}
& \textbf{ABD} 
& \textbf{DO} 
& \textbf{VOID}
& \textbf{VOID} 
& \textbf{Non-VOID} 
& \textbf{ABD} 
& \textbf{DO} 
& \textbf{ABD} 
& \textbf{DO} 
& \textbf{VOID} \\
\hline
Balanced Accuracy (\%) & 83 & 71 & 83 & 84 & 84 & 82 & 82 & 83 & 65 & 84 \\
Sensitivity (\%)       & 78 & 52 & 83 & 90 & 79 & 82 & 83 & 72 & 43 & 90 \\
Specificity (\%)       & 87 & 90 & 84 & 79 & 90 & 83 & 82 & 93 & 87 & 79 \\
\hline
\textbf{Overall Accuracy (\%)} 
& \multicolumn{3}{c|}{74} 
& \multicolumn{2}{c|}{81} 
& \multicolumn{2}{c|}{82} 
& \multicolumn{3}{c|}{70} \\
\hline
\textbf{F1-macro} 
& \multicolumn{3}{c|}{0.68} 
& \multicolumn{2}{c|}{0.76} 
& \multicolumn{2}{c|}{0.75} 
& \multicolumn{3}{c|}{0.64} \\
\hline
\textbf{AUC} 
& \multicolumn{3}{c|}{0.87} 
& \multicolumn{2}{c|}{0.89} 
& \multicolumn{2}{c|}{0.89} 
& \multicolumn{3}{c|}{0.73} \\
\hline
\end{tabular}%
}
\end{table}

% Train on A and B, Test on C

\begin{table}[ht]
\centering
\caption{Performance metrics for Stage 1 (VOID vs Non-VOID), 
Stage 2 (ABD vs DO), and Cascaded Stage 1+2 (ABD, DO, VOID) 
using the MLP classifier, trained on Dataset A and B and tested on Dataset C.}
\label{tab:stage1_stage2_metrics_ABC}
\renewcommand{\arraystretch}{1.3}
\setlength{\extrarowheight}{1pt}
\resizebox{0.49\textwidth}{!}{%
\begin{tabular}{|l|c|c|c|c|c|c|c|c|c|c|}
\hline
\multirow{2}{*}{\textbf{Metric (Tr:A+B, Te:C)}} 
& \multicolumn{3}{c|}{\textbf{Single Stage}}
& \multicolumn{2}{c|}{\textbf{Stage 1}} 
& \multicolumn{2}{c|}{\textbf{Stage 2}} 
& \multicolumn{3}{c|}{\textbf{Cascaded Stage}} \\
\cline{2-11}
& \textbf{ABD} 
& \textbf{DO} 
& \textbf{VOID}
& \textbf{VOID} 
& \textbf{Non-VOID} 
& \textbf{ABD} 
& \textbf{DO} 
& \textbf{ABD} 
& \textbf{DO} 
& \textbf{VOID} \\
\hline
Balanced Accuracy (\%) & 81 & 68 & 70 & 76 & 76 & 80 & 80 & 82 & 68 & 76 \\
Sensitivity (\%)       & 86 & 48 & 51 & 65 & 88 & 94 & 65 & 89 & 43 & 65 \\
Specificity (\%)       & 77 & 89 & 88 & 88 & 65 & 65 & 94 & 75 & 94 & 88 \\
\hline
\textbf{Overall Accuracy (\%)} 
& \multicolumn{3}{c|}{74}
& \multicolumn{2}{c|}{84} 
& \multicolumn{2}{c|}{90} 
& \multicolumn{3}{c|}{77} \\
\hline
\textbf{F1-macro} 
& \multicolumn{3}{c|}{0.61}
& \multicolumn{2}{c|}{0.74} 
& \multicolumn{2}{c|}{0.80} 
& \multicolumn{3}{c|}{0.65} \\
\hline
\textbf{AUC} 
& \multicolumn{3}{c|}{0.83}
& \multicolumn{2}{c|}{0.85} 
& \multicolumn{2}{c|}{0.87} 
& \multicolumn{3}{c|}{0.73} \\
\hline
\end{tabular}%
}
\end{table}

% Train on B, Test on A + C

\begin{table}[ht]
\centering
\caption{Performance metrics for Stage 1 (VOID vs Non-VOID), 
Stage 2 (ABD vs DO), and Cascaded Stage 1+2 (ABD, DO, VOID) 
using the MLP classifier, trained on Dataset B and tested on Dataset A and C.}
\label{tab:stage1_stage2_metrics_BAC}
\renewcommand{\arraystretch}{1.3}
\setlength{\extrarowheight}{1pt}
\resizebox{0.49\textwidth}{!}{%
\begin{tabular}{|l|c|c|c|c|c|c|c|c|c|c|}
\hline
\multirow{2}{*}{\textbf{Metric (Tr:B, Te:A+C)}} 
& \multicolumn{3}{c|}{\textbf{Single Stage}}
& \multicolumn{2}{c|}{\textbf{Stage 1}} 
& \multicolumn{2}{c|}{\textbf{Stage 2}} 
& \multicolumn{3}{c|}{\textbf{Cascaded Stage}} \\
\cline{2-11}
& \textbf{ABD} 
& \textbf{DO} 
& \textbf{VOID}
& \textbf{VOID} 
& \textbf{Non-VOID} 
& \textbf{ABD} 
& \textbf{DO} 
& \textbf{ABD} 
& \textbf{DO} 
& \textbf{VOID} \\
\hline
Balanced Accuracy (\%) & 80 & 63 & 79 & 79 & 79 & 78 & 78 & 78 & 58 & 79 \\
Sensitivity (\%)       & 70 & 41 & 77 & 88 & 70 & 82 & 74 & 64 & 27 & 88 \\
Specificity (\%)       & 89 & 84 & 80 & 70 & 88 & 74 & 82 & 91 & 89 & 70 \\
\hline
\textbf{Overall Accuracy (\%)} 
& \multicolumn{3}{c|}{67}
& \multicolumn{2}{c|}{74} 
& \multicolumn{2}{c|}{81} 
& \multicolumn{3}{c|}{64} \\
\hline
\textbf{F1-macro} 
& \multicolumn{3}{c|}{0.60}
& \multicolumn{2}{c|}{0.71} 
& \multicolumn{2}{c|}{0.67} 
& \multicolumn{3}{c|}{0.56} \\
\hline
\textbf{AUC} 
& \multicolumn{3}{c|}{0.78}
& \multicolumn{2}{c|}{0.82} 
& \multicolumn{2}{c|}{0.80} 
& \multicolumn{3}{c|}{0.62} \\
\hline
\end{tabular}%
}
\end{table}

\begin{figure}[ht]
\centering
\includegraphics[width=0.49\textwidth]{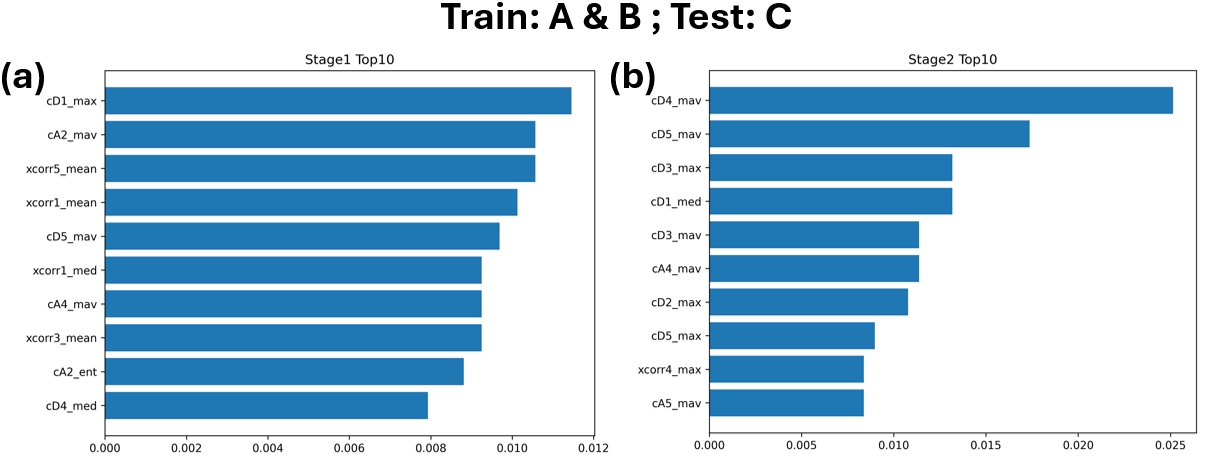}
\caption{Permutation Feature Importance (PFI) for the MLP classifier trained on Datasets A and B and validated on Dataset C. (a) Stage~1 (VOID vs.\ non-VOID) and (b) Stage~2 (ABD vs.\ DO). Bars show the decrease in accuracy after permuting each feature, only the top 10 features are shown. Features include wavelet approximation (cA), detail (cD), and cross-correlation (xcorr) coefficients. PFI for other configurations (random 60/40 split and training on Dataset B with validation on Datasets A and C) is in Supplementary Figure~S2.}

\label{fig:PFI_stage1_stage2}
\end{figure}

In Stage 2, nine of the top ten ranked features were derived from approximation and detail wavelet coefficients (e.g., $cA$, $cD$) when the model was evaluated on Dataset C, while training on Datasets A and B. However, the specific top-ranked features varied when different training and validation datasets were used. This variation likely reflects that multiple computed features contribute meaningfully to distinguishing ABD and DO events, rather than reliance on a small, fixed subset of predictors (Fig.~\ref{fig:PFI_stage1_stage2}, and Figure~S2, panel~a).

%\begin{figure*}[htbp]
%    \centering
%%    \includegraphics[width=0.99\textwidth]{Figures_new/feature_importance_stage1_ANN_sorted.png}
%       \includegraphics[width=0.49\linewidth, height=0.96\textheight, keepaspectratio]{Figures_new/feature_importance_stage1_ANN_sorted.png}
%       \includegraphics[width=0.49\linewidth, height=0.96\textheight, keepaspectratio]{Figures_new/feature_importance_stage2_ANN_sorted.png}
%    \caption{
%        Permutation Feature Importance (PFI) of the ANN classifier for Stage 1 (left panel) and Stage 2 (right panel), showing the drop in accuracy for each feature when permuted. The x-axis represents the PFI (drop in accuracy), and the y-axis lists 55 features including approximation, detail, and cross-correlation between approximation and detail wavelet coefficients.}
%    \label{fig:PFI_stage1_stage2}
%\end{figure*}

\subsection{Cascaded Stage 1+2: Multi-class EVENT Classification (ABD, DO, VOID)}
The cascaded two-stage MLP combines Stage~1 (VOID vs.\ non-VOID detection) and Stage~2 (ABD vs.\ DO classification) to produce a unified three-class output (ABD, DO, VOID). 
%In this configuration, all samples are directly assigned to one of the three classes, leveraging the hierarchical detection-classification strategy in a single cascaded pipeline. 
When externally validated on Dataset C, the cascaded model achieved an overall accuracy of \textbf{77\%} (70\% with 40\% split; 64\% when validated on Datasets A and C) (Tables~\ref{tab:stage1_stage2_metrics_40}, \ref{tab:stage1_stage2_metrics_ABC}, and \ref{tab:stage1_stage2_metrics_BAC}). We achieved strong per-class discrimination for ABD and VOID (Figure~S1), whereas DO classification remained comparatively lower. This reduced performance likely reflects limited DO samples, class imbalance, and inter-dataset variability affecting generalization. ROC analysis demonstrated moderate to strong separability, with AUC values ranging from 59\% to 87\% on Dataset C (Fig.~\ref{fig:all_stages_cm_raw_unseen_test_data}).  

Although the cascaded framework improved performance in the binary subtasks (Stage~1 and Stage~2), overall three-class gains were constrained by error propagation between stages. Nevertheless, the hierarchical design aligns with clinical reasoning and enhances interpretability.

To further assess the robustness of the models across different data splits, additional experiments were conducted using alternative training–-testing configurations, including 80\%–-20\% and 70\%-–30\% random splits, as well as training on dataset B with testing only on dataset C. The results for all stages and configurations are summarized in Tables SI-–SIII in the supplementary material. Across all strategies, the MLP demonstrated stable and consistent performance, indicating its ability to generalize across datasets and maintain robustness under varying data distributions.

\subsection{Visual Interpretation of Detection Results on External Dataset C}
%Figs.~S5--S9 in the Supplementary Material show five representative examples from the test set, illustrating true and predicted event annotations over \Pves\ traces. Each panel corresponds to a different test segment with varying noise levels. The model successfully identified urological events, ABD, DO, VOID, with minimal false positives or missed detections, even in segments with moderate noise levels.
%Fig.~\ref{fig:pves_actual_pred} shows four representative examples from 4 different patients, illustrating actual (A) and predicted (P) event annotations over \Pves\ traces. Each panel corresponds to a \Pves\ trace, from a different patient, with varying noise levels. 

\begin{figure*}[ht]
\centering
\includegraphics[width=0.99\textwidth]{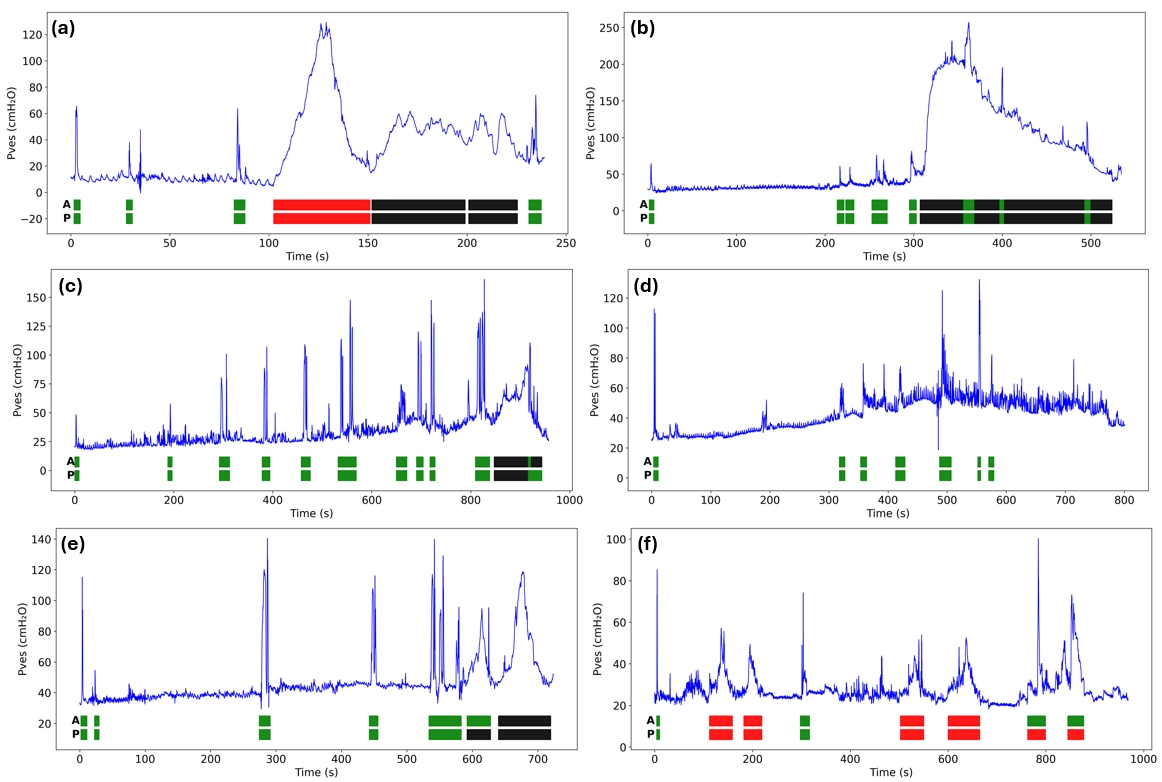}
\caption{Actual (A) and predicted (P) event annotations over bladder pressure (\Pves) traces from six subjects in external validation Dataset C, with varying signal morphologies and noise levels. Panel~a shows accurate detection of all three event types, panel~b highlights correct identification of push (ABD) during voiding, panels~c--d demonstrate robust detection in noisy signals, panels~e--f reveal cases where model predictions differ from manual labels, reflecting inconsistencies in manual annotation rather than model error. Event types are color-coded as follows: abdominal activity (ABD, green), detrusor overactivity (DO, red), and voiding contractions (VOID, black). The model was trained on Datasets A and B.}
\label{fig:pves_actual_pred}
\end{figure*}

The model accurately detected urological events (ABD, DO, VOID) with minimal false positives, even in moderately noisy signals (Fig.~\ref{fig:pves_actual_pred}). Dataset C was used for external validation after training on Datasets A and B. The model also identified push (abdominal activity, ABD) during voiding (VOID), seen in Panels~b and~c, a clinically relevant phenomenon.

Panels~e and~f highlight the cases where the predicted events differ from the manual annotations: in Panel~e, the model predicts a VOID event while the manual annotation lists ABD, but the morphology is more consistent with VOID, as supported by the subsequent event. Similarly, in Panel~f, the last event appears morphologically more similar to DO than the annotated ABD. These examples are included to illustrate that apparent model errors may instead reflect limitations or inconsistencies in manual labeling. These differences likely reflect limitations of manual annotation rather than model error, as subtle events can be missed even by experienced urologists due to known interobserver variability in urodynamic interpretation \cite{dudley_interrater_2017,dudley_interrater_2018}, and visual inspection shows that the model detects them accurately.

Additional examples from four other subjects are provided in Figure~S1 (Supplementary Material), confirming consistent detection across diverse signal conditions. Together, these figures provide a comprehensive visual assessment of the model's detection performance and help justify the reliability of predictions, even when manual annotations are imperfect. From these 10 \Pves\ traces, only two contain DO events, clearly illustrating that the reduced performance of DO (Table~\ref{tab:stage1_stage2_metrics_ABC}) compared to ABD and VOID likely reflects the limited number of DO samples in external Dataset C rather than model inadequacy.

\subsection{Proof-of-Concept Evaluation on UM Ambulatory Data}
%Fig.~\ref{fig:pves_UM_pred} presents a representative example from the UM external validation dataset. 

\begin{figure*}[!htb]
    \centering
    \includegraphics[width=0.85\textwidth]{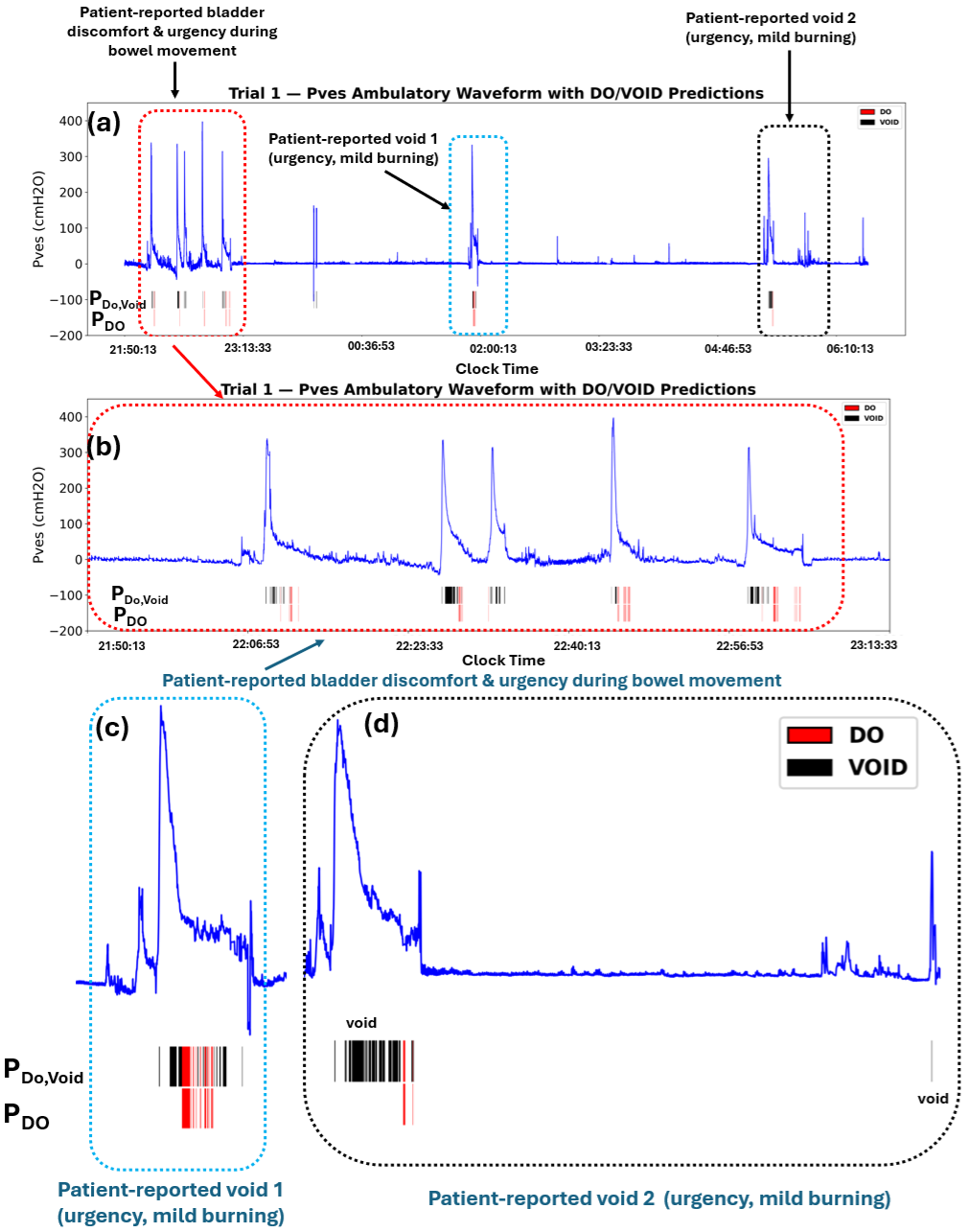} %B
    \caption{Example of proof-of-concept evaluation using ambulatory \Pves\ recordings from the UM device.  
Panels a--d show representative \Pves\ recordings with patient-reported symptoms indicated by black/blue arrows/text and model predictions as $P_{DO,Void}$ restricted to DO and VOID, and $P_{DO}$ restricted to DO labels. Dotted boxes of the same color indicate corresponding regions in the full trace and their zoomed-in views. Color-coded label bars denote event types: red for DO, and black for VOID. The model was applied directly to raw \Pves\ signals, enabling visual assessment of its ability to detect physiologically relevant urological events in an external dataset.
}
    \label{fig:pves_UM_pred}
\end{figure*}

Despite differences in device characteristics, variable signal quality, and the absence of any signal preprocessing, the model accurately identified most clinically relevant urological events such as DO and VOID, in ambulatory data using our UM system (Fig.~\ref{fig:pves_UM_pred}). Most DO and VOID events were correctly detected in the predicted labels, consistent with patient-reported symptoms, demonstrating the model's generalizability to external, real-world data.

Notably, the bottom label bars (DO-only, $P_{DO}$) emphasize DO activity in this patient. The vertical predicted labels show that the two-stage predictions, trained on Dataset A and Dataset B, successfully captured DO, and VOID events (Fig.~\ref{fig:pves_UM_pred}). 

These results illustrate that that the trained two-stage MLP model, based on Dataset A and Dataset B, has the potential to retain meaningful discriminatory capability when applied to raw ambulatory \Pves\ signals.

%Additionally, we compared our results with previously published studies on urological event detection,  acknowledging that these used different datasets and thus serve only as indicative performance references \hassaan{here, I will add a table by comparing our results with previous study/studies later, if required}.

%Figure 1, confusion matrix and ROC curve using ANN, KNN, SVM, Random Forest and Decision tree (using single stage classifier)
% Table 1 shows the performance for each classifier
%Figure 2, includes the confusion matrix:single vs 2-stage ANN classifier and with its performance mertrix in table 2

\section{Discussion}
\label{discussion}
Our proposed two-stage classification framework automatically detects urological events from \Pves\ alone and demonstrated consistent performance across multiple evaluation settings, including independent and cross-dataset validation. Stage~1 (VOID vs.\ non-VOID) achieved accuracies ranging from 74\% to 84\% with AUC values up to 0.89, while Stage~2 (ABD vs.\ DO) achieved accuracies between 81\% and 90\% with AUC values up to 0.89 across evaluation settings. Balanced accuracy remained relatively stable despite class imbalance, particularly for the less frequent DO events. A proof-of-concept evaluation on ambulatory data further demonstrated that most clinically relevant events were successfully detected. Permutation feature importance analysis indicated that the majority of the analyzed wavelet coefficient–-based features contributed meaningfully to classification performance. Together, these findings highlight the potential of AI-driven approaches to improve the objective and real-time interpretation of urodynamic studies.

%\subsection{Confusion Matrix Analysis}
%The confusion matrix for each event type further support the model’s ability to distinguish between closely related urological events with high specificity. Only a few false positives were observed, primarily due to overlapping signal characteristics between detrusor overactivity and bladder contractions.

\subsection{Model Interpretation and Comparison with Prior Studies}
The consistent performance of the two-stage MLP suggests that wavelet-transformed features effectively capture the key time–-frequency characteristics of bladder pressure signals. As the framework relies on engineered statistical and wavelet-based features rather than raw time-series data, a fully connected MLP is well suited to learn nonlinear relationships while maintaining moderate model complexity. This feature-based design also enables interpretable analysis via permutation feature importance, highlighting the contribution of multiple wavelet-derived features to classification performance.

%The strong and consistent performance of the two-stage MLP across evaluation settings suggests that the wavelet--transformed features effectively capture the key time–-frequency characteristics of bladder pressure signals. The MLP model was selected due to its practical advantages for real-time hardware implementation. Feedforward MLP architectures involve relatively simple computations \cite{dong_implementation_2006}, making them suitable for deployment on embedded platforms such as FPGAs or microcontrollers, with low inference latency and power consumption \cite{magdy_saady_hardware_2022,faris_ali_efficient_2023,nguyen_efficient_2018,botros_hardware_1993}. Recurrent models like LSTM and BiLSTM, while strong in sequential data modeling, introduce higher computational complexity, which may limit efficiency in low-power or embedded systems. Similarly, tree-based models like Random Forests are less commonly optimized for hardware acceleration. Thus, MLP provides an optimal balance between accuracy, efficiency, and implementability, supporting its use in real-time clinical decision-making \cite{zafeiriou_comparative_2024}.  

Previous work in automated urodynamics has largely focused on signal thresholding, rule-based approaches, or statistical methods \cite{karam_real-time_2016}. Our results extend earlier work by Abbaraju et al., who first proposed automated classification from single-channel \Pves\ signals using machine learning \cite{abbaraju_real-time_2023}. While their framework demonstrated feasibility, performance was reduced in ambiguous or artifact-prone segments. By implementing a two-stage MLP architecture, our study addresses these limitations, achieving improved event discrimination, enhanced resilience to signal artifacts, and robustness to signal variability. Finally, we strengthened robustness and clinical applicability through rigorous train-–test splitting, randomized cross-validation, and external validation on independent datasets.

Previous reviews have summarized the application of machine learning in urodynamics, highlighting its potential for improving feature extraction, diagnosis, and outcome prediction, while noting limitations such as retrospective, single-center studies, limited external validation, and sensitivity to noise \cite{liu_applications_2024, gammie_can_2024}. These findings underscore the need for robust, generalizable models validated across diverse datasets, a gap our study addresses through a two-stage MLP framework evaluated on both controlled and ambulatory datasets, from three different centers.

In other studies, machine learning was applied to detect only one type of urological event, such as DO \cite{wang_pattern_2021,hobbs_machine_2022}, and typically relied on multi-channel inputs. In contrast, our study demonstrates the feasibility of detecting three distinct events, VOID, DO and ABD, from only \Pves\ signals. This approach may pave the way for future applications in ambulatory urodynamic monitoring. Moreover, our multi-center dataset, which includes both male and female participants across a wide age range, further supports the robustness and generalizability of our findings.

Table~\ref{tab:prior_work_comparison} summarizes prior studies on automated urological event detection and positions our proposed two-stage framework in this context. Earlier studies primarily focused on detecting DO or bladder contractions using multiple channels (typically 3), applying handcrafted features with rule-based, thresholding, or traditional ML approaches such as dynamic time warping \cite{wang_pattern_2021}, SVM \cite{hobbs_machine_2022}, or FFT-based analysis \cite{cullingsworth_automated_2018}. While these approaches achieved moderate accuracy (e.g., 81--92\% AUC or 77--84\% sensitivity), they were limited to binary classification and may suffer in the presence of overlapping events. More recent work has applied deep learning models, including CNNs and wavelet-based hybrid approaches \cite{zhou_pilot_2023, batista_miranda_2025}, achieving higher accuracy. However, these models were still primarily focused on DO detection and relied on multi-channel signals, limiting their generalizability to real-world clinical settings. 
%Only a few studies \cite{abbaraju_real-time_2023, majerus_real-time_2024} attempted multi-class classification using single-channel \Pves\ data, reporting moderate performance (overall accuracy ~89\% on cross-validation), highlighting the challenge of robust multi-class prediction from minimal input.

In contrast, our two-stage framework demonstrated strong performance on independent external Dataset C, achieving overall accuracy of 84\% in Stage 1 and 90\% in Stage 2, with AUC values of 0.85 and 0.87, and balanced accuracies of 76\% and 80\%, respectively. Importantly, the model successfully detects ABD, DO, and VOID events from \Pves\ alone, maintaining robust prediction and generalizability even in challenging or noisy signals.

\begin{table*}[htbp]
\caption{Comparison of the proposed two-stage classification framework with prior studies on automated urological event detection.}
\label{tab:prior_work_comparison}
\centering
\renewcommand{\arraystretch}{1.2} 
\begin{tabularx}{\textwidth}{c c c c c c c c X} 
\toprule
\textbf{Work} & \textbf{Year} & \textbf{Application} & \textbf{Channels} & \textbf{Feature} & \textbf{Classes} & \textbf{Classifier} & \textbf{Findings} \\
\midrule
\makecell{Wang et al.\\\cite{wang_pattern_2021}} & 2020 & \makecell{DO\\Detection} & 3 & Wave-shape manifold & 2 & \makecell{Dynamic\\time\\ warping} & \makecell{Acc. 81\% \\ ROC AUC 0.84 \\ Sens. 77\% \\ Spec. 81\%} \\\addlinespace[3mm]

\makecell{Hobbs et al.\\\cite{hobbs_machine_2022}} & 2021 & \makecell{DO\\Detection} & 3 & Windowed time/freq. & 2 & SVM (RBF) & \makecell{ROC AUC 0.92 \\ Sens. 84\% \\ Spec. 93\%} \\\addlinespace[3mm]

\makecell{Karam et al.\\\cite{karam_real-time_2016}} & 2016 & \makecell{Bladder\\Contractions} & 1 & 5--level DWT & 2 & CAT & \makecell{TP 97\%} \\\addlinespace[3mm]

\makecell{Cullingsworth et al.\\\cite{cullingsworth_automated_2018}} & 2018 & \makecell{DO\\Detection} & 3 & Largest rhythmic peaks & 2 & FFT & \makecell{Sens. 27\% \\ Spec. 100\%} \\\addlinespace[3mm]

\makecell{Zhou et al.\\\cite{zhou_pilot_2023}} & 2023 & \makecell{DO\\Detection} & 3 & Threshold screening & 2 & CNN & \makecell{Acc. 100\%} \\\addlinespace[3mm]

\makecell{Batista-Miranda et al.\\\cite{batista_miranda_2025}} & 2025 
& \makecell{DO\\Detection} & 3  & \makecell{DWT\\(Wavelet-based)}  & 2 
& \makecell{CNN +\\Wavelet ML}  & \makecell{Acc. 84.2\% \\ Sens. 82.6\% \\ Spec. 82.6\%} \\\addlinespace[3mm]

%\makecell{Abbaraju et al.  \\\cite{abbaraju_real-time_2023} \\ Majerus et al. \\\cite{majerus_real-time_2024}}  & 2023/2024 & \makecell{ABD, DO,\\VOID, NONE} & 1 & 5--level DWT & 4 & \makecell{ANN, KNN,\\SVM} & \makecell{ANN (CV only): \\ Spec. 96\% \\ Sens. 90--92\% \\ Prec. 89\% \\ F1 89--91\% \\ ROC AUC 89--99\% \\ Overall.Acc 89\%} \\\addlinespace[3mm]

%\makecell{Our Single-Stage \\ Refined, enhanced, and \\ extended from \cite{abbaraju_real-time_2023,majerus_real-time_2024}} & 2025 & \makecell{ABD, DO,\\VOID} & 1 & 5--level DWT & 3 & \makecell{ANN} & \makecell{ANN (unseen test set): \\ Spec. 91--99\% \\ Sens. 84--94\% \\ Prec. 54--96\% \\ F1 68--92\% \\ ROC AUC 98--99\% \\ Overall.Acc 88\%} \\\addlinespace[3mm]

\makecell{\textbf{Our Proposed (Two-Stage)} \\ Refined, enhanced, and \\ extended from \cite{abbaraju_real-time_2023,majerus_real-time_2024}} & 2026 & \makecell{ABD, DO,\\VOID} & 1 & 5--level DWT & 3 & \makecell{MLP} & \makecell{MLP (externally validated: Dataset C): \\ Bal.Acc. 76\%(s1), 80\%(s2), 68--82\%(cascaded) \\Acc. 84\%(s1), 90\%(s2), 77\%(cascaded) \\ F1-macro 74\%(s1), 80\%(s2), 65\%(cascaded) \\ AUC. 85\%(s1), 87\%(s2), 73\%(cascaded)}
\\
\bottomrule
\end{tabularx}
\end{table*}

\subsection{Proof-of-Concept Evaluation on UM Ambulatory Data}
The preliminary proof-of-concept evaluation on ambulatory data from a single subject demonstrates that the proposed two-stage MLP-based detection framework has the potential to generalize to ambulatory \Pves\ recordings acquired from a different device and in a different environment.  These results demonstrate that the model can capture clinically validated urological occurrences directly from raw ambulatory \Pves\ recordings, suggesting that it identifies most of the meaningful signal signatures rather than overfitting to device-specific artifacts. 

A key limitation is the absence of ground-truth annotations in the ambulatory \Pves\ dataset, which prevents formal quantification of detection performance. Ongoing efforts aim to expand the dataset and obtain expert-labeled events to enable rigorous evaluation of sensitivity, specificity, and event-level accuracy. %Future work will also explore domain adaptation and noise-robust preprocessing strategies to further enhance model applicability in real-world ambulatory settings. 
Overall, these results support the potential of the system to assist in the interpretation of data from continuous ambulatory bladder monitoring outside the laboratory environment.

\subsection{Clinical Implications}
%Current clinical practice guidelines, such as the American Urological Association (AUA)/Society of Urodynamics, Female Pelvic Medicine \& Urogenital Reconstruction (SUFU), on the diagnosis and treatment of Idiopathic Overactive Bladder (OAB), emphasize symptom questionnaires, voiding diaries, and urinalysis for clinical evaluation \cite{cameron_aua_2024}. While these approaches provide valuable diagnostic information, they are largely subjective and limited in capturing the underlying physiological dynamics of bladder function. Urodynamic assessment, including the detection of urological events such as ABD, DO, and VOID from \Pves, offers an objective measure of bladder activity. 
UDS are commonly used in clinical settings to evaluate lower urinary tract function and assist in the diagnosis of conditions such as idiopathic OAB \cite{cameron_aua_2024}. UDS provides objective information on bladder activity, including detection of urological events such as ABD, DO, and VOID from \Pves. However, the invasive and clinic-based nature of UDS may introduce patient burden and may not fully capture symptom patterns experienced in everyday settings \cite{finkelstein_anxiety_2020,oktaviani_intraurethral_2021,vogt_catheter-free_2024}.

Ambulatory urodynamic monitoring provides a more physiologically representative assessment of bladder function by enabling natural filling during daily activities, but it also introduces significant challenges for data interpretation \cite{chew_future_2025}. Compared with conventional laboratory-based urodynamics, ambulatory recordings are longer in duration and more susceptible to motion artifacts, posture changes, and variable external conditions, increasing signal variability and complicating manual event annotation \cite{moeyersons_artefact_2019,chew_future_2025}. Prior reviews have noted that although ambulatory approaches may improve sensitivity to underlying pathophysiology, movement-related artifacts and noise remain key limitations \cite{moeyersons_artefact_2019,chew_future_2025}. Recent work on wireless and catheter-free ambulatory bladder monitoring further highlights that prolonged pressure recordings can contain substantial physiological noise from motion and sudden movements, which may lead to false positives in automated event detection if not properly addressed \cite{zareen_optimization_2024}.

%In this context, the proposed ANN-based framework demonstrates the potential to address these challenges by enabling automated detection of ABD, DO, and VOID events from noisy \Pves\ signals using a single channel. While the model was primarily developed and externally validated on conventional urodynamic studies, a proof-of-concept evaluation on limited ambulatory data suggests its feasibility for interpreting prolonged, noise-contaminated recordings. By simplifying instrumentation while maintaining diagnostic accuracy, the framework could support portable or point-of-care urodynamic systems, expand access in lower-resource settings, and facilitate integration with telemedicine platforms \cite{chew_future_2025}. Incorporating our ANN event detection approach could enhance diagnostic precision by providing quantitative physiological measures that complement symptom-based OAB evaluation, supporting more objective and patient-specific assessment. Several studies have also demonstrated the feasibility of AI-assisted approaches in functional urology, highlighting their potential to support clinicians in real-time assessment and personalized care \cite{bentellis_artificial_2021,gammie_can_2024,liu_applications_2024,wang_pattern_2021,hobbs_machine_2022}. These findings suggest a pathway toward streamlined and automated urodynamic interpretation, with future studies needed to validate performance in larger ambulatory cohorts.

In this context, the proposed MLP-based framework demonstrates the potential to address these challenges by enabling automated detection of ABD, DO, and VOID events from noisy \Pves\ signals using only \Pves\ data. The model was primarily developed and externally validated on conventional urodynamic studies, and a proof-of-concept evaluation on limited ambulatory data suggests its feasibility for interpreting prolonged, noise-contaminated recordings. Incorporating this MLP event detection approach could enhance diagnostic precision by providing quantitative physiological measures that complement symptom-based OAB evaluation, supporting more objective and patient-specific assessment. These findings indicate a pathway toward streamlined and automated urodynamic interpretation, with future studies needed to validate performance in larger ambulatory cohorts.

%Accurate identification of ABD, DO, and VOID events is critical for clinical decision-making during urodynamic testing. Misinterpretation can lead to inappropriate diagnoses, such as overdiagnosing detrusor overactivity, and suboptimal treatment selection. 

%In this study, the proposed single-channel vesical pressure framework simplifies instrumentation while maintaining diagnostic accuracy, creating the potential for portable or point-of-care urodynamic systems. Such applications could expand access in lower-resource settings and support telemedicine integration \cite{chew_future_2025}. Moreover, several studies have demonstrated the feasibility of AI-assisted approaches in functional urology, highlighting their potential to support clinicians in real-time assessment and personalized care \cite{bentellis_artificial_2021,gammie_can_2024,liu_applications_2024,wang_pattern_2021,hobbs_machine_2022}.

\subsection{Study limitations and future research}
This study analyzed 118 UDS traces from 76 patients. Future work should evaluate the proposed methods on larger and more diverse cohorts, incorporate manual annotations from multiple expert urologists, and quantitatively assess ambulatory performance by comparing model predictions against ground-truth labels. 

%The current analysis focused on three event types (ABD, DO, and VOID), excluding other clinically relevant lower urinary tract events \cite{bang_feasibility_2022, tsai_building_2024}. Extending the model from single-label to multi-label classification would further enable detection of overlapping events (e.g., abdominal straining such as push during voiding).

The model was externally validated on independent datasets from another center, demonstrating cross-center generalizability. However, prospective real-time evaluation remains necessary to fully assess robustness under clinical deployment conditions. Ongoing data collection will support such validation.

%Future work should explore additional models, including convolutional neural networks (CNN), long short-term memory networks (LSTM), support vector machines (SVM), and other classifiers, and compare their performance with the current MLP model.

\section{Conclusions}
\label{conclusions}

Accurate detection of urological events is essential for optimizing the diagnosis and management of bladder disorders. We demonstrated that a two-stage MLP can reliably detect ABD, DO, and VOID events from single-channel \Pves\ signals, maintaining high accuracy and specificity across both internal testing and external independent validation datasets, marking an important step toward practical AI-assisted urodynamics. By refining previous approaches, our framework addresses key limitations in earlier methods. PFI analysis further indicates that most wavelet-based features contribute meaningfully to model performance. Finally, a proof-of-concept evaluation on ambulatory \Pves\ data demonstrates the potential of the model to assist in interpreting noisy, prolonged ambulatory recordings.

\section*{Acknowledgment}
This research was supported by funding from the Glickman Urology Institute at the Cleveland Clinic, the U.S. Department of Veterans Affairs (Grant RX003687), the Advanced Platform Technology (APT) Center, and Case Western Reserve University. The authors also thank the patients who participated in the studies, as well as the expert urologists and clinical staff at the Louis Stokes Cleveland VA Medical Center, the Cleveland Clinic, and the University of Pittsburgh for their valuable contributions to data collection and annotation.

%ask Steve and Becky for funding support to add here

\bibliography{refs_AI}
%\bibliographystyle{BibLaTeX}
%\bibliographystyle{ieeetr}

%\ifCLASSOPTIONcaptionsoff
%  \newpage
%\fi




\bibliographystyle{IEEEtran}

\end{document}

\bstctlcite{IEEEexample:BSTcontrol}

%\linenumbers   % after \begin{document} for line numbers

\title{
Supplementary Information: Automated Detection of Urological Events in Bladder Pressure Signals with a Two-Stage Machine Learning Framework Validated on External Datasets
}

\author{Hassaan A. Bukhari$^{1}$, 
    Vikram Abbaraju$^{2}$, 
    Jay Patel$^{3,4}$, 
    Becky Clarkson$^{5}$, 
    Shachi Tyagi$^{5}$, 
    Margot S. Damaser$^{1,4,6}$,
    Steve J. A. Majerus$^{3,4}$ \\
\ \\ % leave an empty line between authors and affiliation
$^1$ Department of Biomedical Engineering, Cleveland Clinic Research, Cleveland, OH, USA\\
$^2$ School of Electrical and Computer Engineering, Georgia Institute of Technology, Atlanta, GA, USA\\
$^3$ Case Western Reserve University, Cleveland, OH, USA\\
$^4$ Advanced Platform Technology Center, Louis Stokes Cleveland Veterans Affairs Medical Center, Cleveland, OH, USA \\
$^5$ Division of Geriatric Medicine, University of Pittsburgh, Pittsburgh, PA, USA \\
$^6$ Glickman Urology Institute, Cleveland Clinic, Cleveland, OH, USA\\}

\markboth{Engineering Applications of Artificial Intelligence}%
%\markboth{manuscript in preparation}%
{Bukhari \MakeLowercase{\textit{et al.}}: Automated Detection of Urological Events}

\maketitle

\clearpage
\onecolumn

\section*{Supplementary Material}

%\begin{figure*}[ht]
%\centering
%\includegraphics[width=1\textwidth]{Figures_new/stage1_confusion_matrix_counts_percent_v8.png}
%%\includegraphics[width=0.49\textwidth]{Figures_new/stage1_confusion_matrix_percent_relabel_data_v3.png}
%\caption{Confusion matrix for Stage 1 classification (raw counts in the left panel and normalized percentages in the right panel) obtained from five-fold cross-validation on the training data. These results show the model’s performance during cross-validation and do not represent evaluation on the independent, unseen test set.}
%\label{fig:stage1_cm_cm_percent}
%\end{figure*}

%\begin{figure*}[ht]
%\centering
%\includegraphics[width=1\textwidth]{Figures_new/stage2_confusion_matrix_counts_percent_v8.png}
%%\includegraphics[width=0.49\textwidth]{Figures_new/stage2_confusion_matrix_percent_relabel_data_v3.png}
%\caption{Confusion matrix for Stage 2 classification (raw counts in the left panel and normalized percentages in the right panel) obtained from five-fold cross-validation on the training data. These results illustrate classifier performance during cross-validation and do not reflect performance on the independent, unseen test set.}
%\label{fig:stage2_cm_raw}
%\end{figure*}

%20% unseen dataset

\begin{table*}[h]
\centering
\caption{Performance metrics for Single Stage (ABD, DO, VOID), Stage 1 (VOID vs Non-VOID), 
Stage 2 (ABD vs DO), and Cascaded Stage 1+2 (ABD, DO, VOID) 
using the ANN classifier on the 20\% unseen test set (Train:80\%, Test:20\%).}
\label{tab:stage1_stage2_metrics_20}
\resizebox{\textwidth}{!}{%
\begin{tabular}{|l|c|c|c|c|c|c|c|c|c|c|}
\hline
\multirow{2}{*}{\textbf{Metric(Tr:80\%,Te:20\%)}} 
& \multicolumn{3}{c|}{\textbf{Single Stage}}
& \multicolumn{2}{c|}{\textbf{Stage 1}} 
& \multicolumn{2}{c|}{\textbf{Stage 2}} 
& \multicolumn{3}{c|}{\textbf{Cascaded Stage}} \\
\cline{2-11}
& \textbf{ABD} 
& \textbf{DO} 
& \textbf{VOID}
& \textbf{VOID} 
& \textbf{Non-VOID} 
& \textbf{ABD} 
& \textbf{DO} 
& \textbf{ABD} 
& \textbf{DO} 
& \textbf{VOID} \\
\hline
Balanced Accuracy (\%) & 82 & 77 & 83 & 86 & 86 & 73 & 73 & 76 & 69 & 86 \\
Sensitivity (\%)       & 79 & 67 & 77 & 86 & 85 & 86 & 59 & 75 & 48 & 86 \\
Specificity (\%)       & 86 & 88 & 89 & 85 & 86 & 59 & 86 & 78 & 90 & 85 \\
\hline
\textbf{Overall Accuracy (\%)} 
& \multicolumn{3}{c|}{0.76}
& \multicolumn{2}{c|}{85} 
& \multicolumn{2}{c|}{79} 
& \multicolumn{3}{c|}{71} \\
\hline
\textbf{F1-macro} 
& \multicolumn{3}{c|}{0.72}
& \multicolumn{2}{c|}{0.80} 
& \multicolumn{2}{c|}{0.73} 
& \multicolumn{3}{c|}{0.67} \\
\hline
\textbf{AUC} 
& \multicolumn{3}{c|}{0.87}
& \multicolumn{2}{c|}{0.89} 
& \multicolumn{2}{c|}{0.84} 
& \multicolumn{3}{c|}{0.74} \\
\hline
\end{tabular}%
}
\end{table*}

%30% unseen dataset

\begin{table*}[h]
\centering
\caption{Performance metrics for Single Stage (ABD, DO, VOID), Stage 1 (VOID vs Non-VOID), 
Stage 2 (ABD vs DO), and Cascaded Stage 1+2 (ABD, DO, VOID) 
using the ANN classifier on the 30\% unseen test set (Train:70\%, Test:30\%).}
\label{tab:stage1_stage2_metrics_30}
\resizebox{\textwidth}{!}{%
\begin{tabular}{|l|c|c|c|c|c|c|c|c|c|c|}
\hline
\multirow{2}{*}{\textbf{Metric(Tr:70\%,Te:30\%)}} 
& \multicolumn{3}{c|}{\textbf{Single Stage}}
& \multicolumn{2}{c|}{\textbf{Stage 1}} 
& \multicolumn{2}{c|}{\textbf{Stage 2}} 
& \multicolumn{3}{c|}{\textbf{Cascaded Stage}} \\
\cline{2-11}
& \textbf{ABD} 
& \textbf{DO} 
& \textbf{VOID}
& \textbf{VOID} 
& \textbf{Non-VOID} 
& \textbf{ABD} 
& \textbf{DO} 
& \textbf{ABD} 
& \textbf{DO} 
& \textbf{VOID} \\
\hline
Balanced Accuracy (\%) & 82 & 64 & 87 & 85 & 85 & 79 & 79 & 83 & 64 & 85 \\
Sensitivity (\%)       & 76 & 39 & 94 & 94 & 77 & 89 & 69 & 75 & 35 & 94 \\
Specificity (\%)       & 89 & 90 & 81 & 77 & 94 & 69 & 89 & 90 & 92 & 77 \\
\hline
\textbf{Overall Accuracy (\%)} 
& \multicolumn{3}{c|}{0.72}
& \multicolumn{2}{c|}{80} 
& \multicolumn{2}{c|}{86} 
& \multicolumn{3}{c|}{71} \\
\hline
\textbf{F1-macro} 
& \multicolumn{3}{c|}{0.65}
& \multicolumn{2}{c|}{0.75} 
& \multicolumn{2}{c|}{0.76} 
& \multicolumn{3}{c|}{0.63} \\
\hline
\textbf{AUC} 
& \multicolumn{3}{c|}{0.86}
& \multicolumn{2}{c|}{0.91} 
& \multicolumn{2}{c|}{0.86} 
& \multicolumn{3}{c|}{0.75} \\
\hline
\end{tabular}%
}
\end{table*}

% Train on B, Test on C

\begin{table*}[h]
\centering
\caption{Performance metrics for Single Stage (ABD, DO, VOID), Stage 1 (VOID vs Non-VOID), 
Stage 2 (ABD vs DO), and Cascaded Stage 1+2 (ABD, DO, VOID) 
using the ANN classifier, trained on Dataset B and tested on Dataset C.}
\label{tab:stage1_stage2_metrics_BC}
\resizebox{\textwidth}{!}{%
\begin{tabular}{|l|c|c|c|c|c|c|c|c|c|c|}
\hline
\multirow{2}{*}{\textbf{Metric (Tr:B, Te:C)}} 
& \multicolumn{3}{c|}{\textbf{Single Stage}}
& \multicolumn{2}{c|}{\textbf{Stage 1}} 
& \multicolumn{2}{c|}{\textbf{Stage 2}} 
& \multicolumn{3}{c|}{\textbf{Cascaded Stage}} \\
\cline{2-11}
& \textbf{ABD} 
& \textbf{DO} 
& \textbf{VOID}
& \textbf{VOID} 
& \textbf{Non-VOID} 
& \textbf{ABD} 
& \textbf{DO} 
& \textbf{ABD} 
& \textbf{DO} 
& \textbf{VOID} \\
\hline
Balanced Accuracy (\%) & 84 & 64 & 75 & 78 & 78 & 77 & 77 & 82 & 59 & 78 \\
Sensitivity (\%)       & 88 & 38 & 62 & 76 & 80 & 90 & 65 & 81 & 28 & 76 \\
Specificity (\%)       & 81 & 90 & 87 & 80 & 76 & 65 & 90 & 83 & 91 & 80 \\
\hline
\textbf{Overall Accuracy (\%)} 
& \multicolumn{3}{c|}{0.75}
& \multicolumn{2}{c|}{79} 
& \multicolumn{2}{c|}{87} 
& \multicolumn{3}{c|}{70} \\
\hline
\textbf{F1-macro} 
& \multicolumn{3}{c|}{0.62}
& \multicolumn{2}{c|}{0.70} 
& \multicolumn{2}{c|}{0.72} 
& \multicolumn{3}{c|}{0.57} \\
\hline
\textbf{AUC} 
& \multicolumn{3}{c|}{0.81}
& \multicolumn{2}{c|}{0.83} 
& \multicolumn{2}{c|}{0.79} 
& \multicolumn{3}{c|}{0.65} \\
\hline
\end{tabular}%
}
\end{table*}

\begin{figure*}[ht]
\centering
\includegraphics[width=\textwidth]{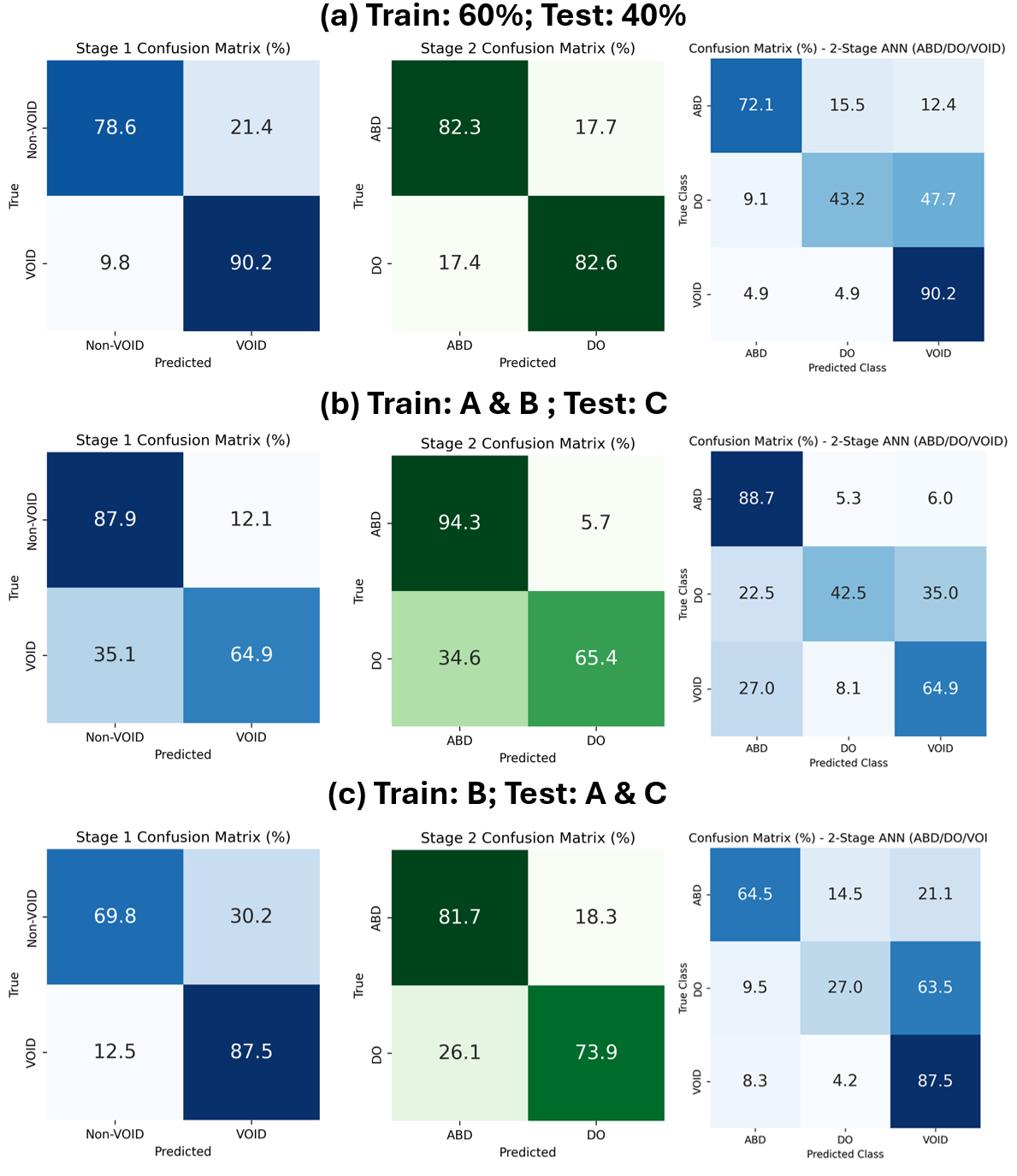}
\caption{Normalized confusion matrices (\%) for different training and validation configurations. 
Panel (a) presents results obtained using a random split (60\% training, 40\% testing). 
Panel (b) shows performance when the model was trained on Datasets A and B and externally validated on Dataset C. 
Panel (c) illustrates results when trained on Dataset B and externally validated on Datasets A and C. 
Within each panel, the left sub-panel corresponds to Stage~1 (VOID vs.\ non-VOID), the middle sub-panel to Stage~2 (ABD vs.\ DO among predicted non-VOID samples), and the right sub-panel to the cascaded two-stage ANN producing the final three-class output.}
\label{fig:all_stages_cm_raw_unseen_test_data}
\end{figure*}

\begin{figure*}[ht]
\centering
\includegraphics[width=\textwidth]{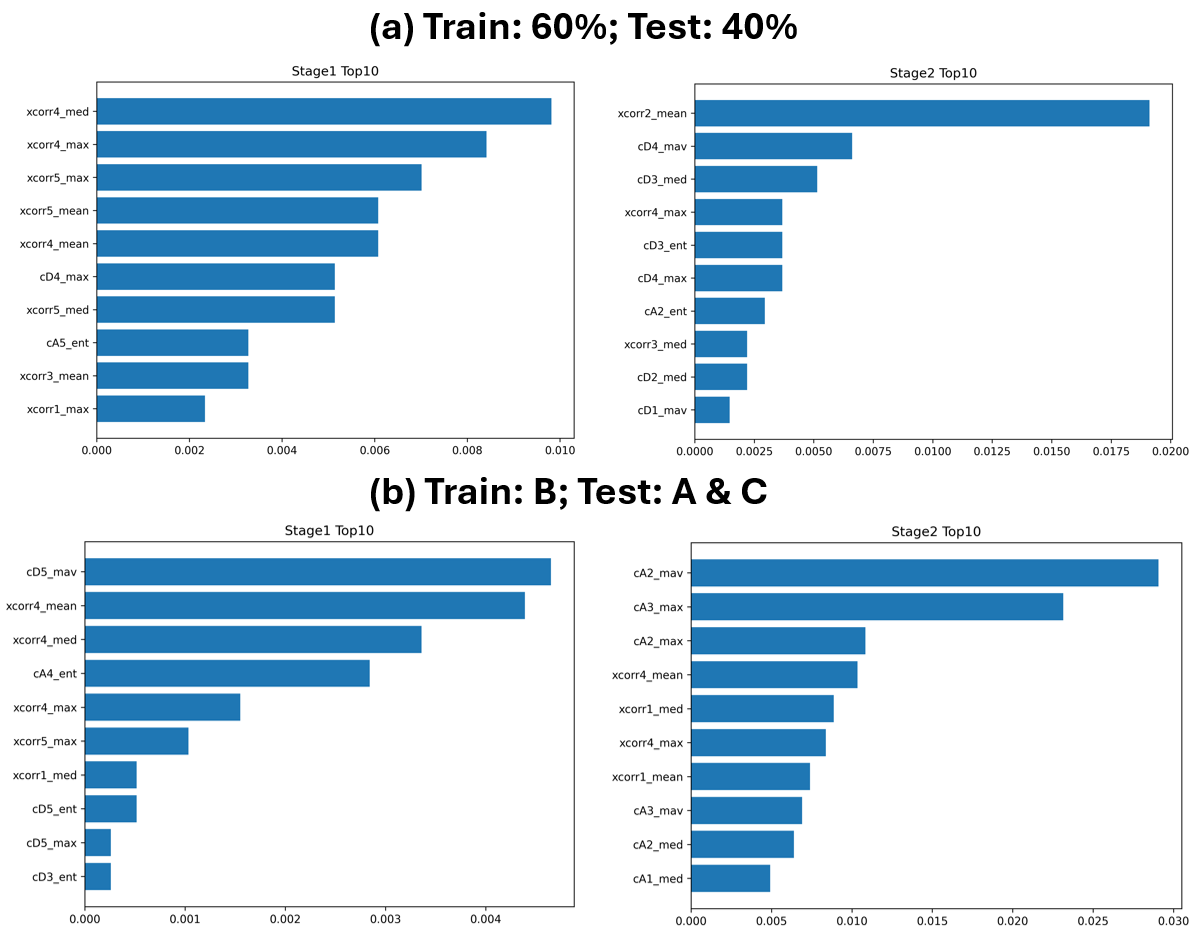}
\caption{Permutation Feature Importance (PFI) results for additional experimental configurations. The top panel (a) shows results obtained using a random 60\%/40\% train-–test split, while the bottom panel (b) presents results when the model was trained on Dataset B and externally validated on Datasets A and C. In each panel, the left sub-panel corresponds to Stage~1 (VOID vs.\ non-VOID) and the right sub-panel to Stage~2 (ABD vs.\ DO). Bars represent the decrease in model accuracy after permuting each feature. For clarity, only the top 10 features are displayed in each sub-panel.}
\label{fig:PFI_stage1_stage2}
\end{figure*}

\begin{figure*}[h!]
    \centering

    % ---- Row 1 ----
   % \begin{subfigure}[t]{0.32\textwidth}
   %     \centering
     %   \includegraphics[width=\textwidth]{Figures_new/trial_064_onepanel.png}
     %   \caption{Patient 1}
   % \end{subfigure}
   % \begin{subfigure}[t]{0.32\textwidth}
   %     \centering
   %     \includegraphics[width=\textwidth]{Figures_new/trial_010_onepanel.png}
   %     \caption{Patient 2}
   % \end{subfigure}
    \begin{subfigure}[t]{0.49\textwidth}
        \centering
                \includegraphics[width=\textwidth]{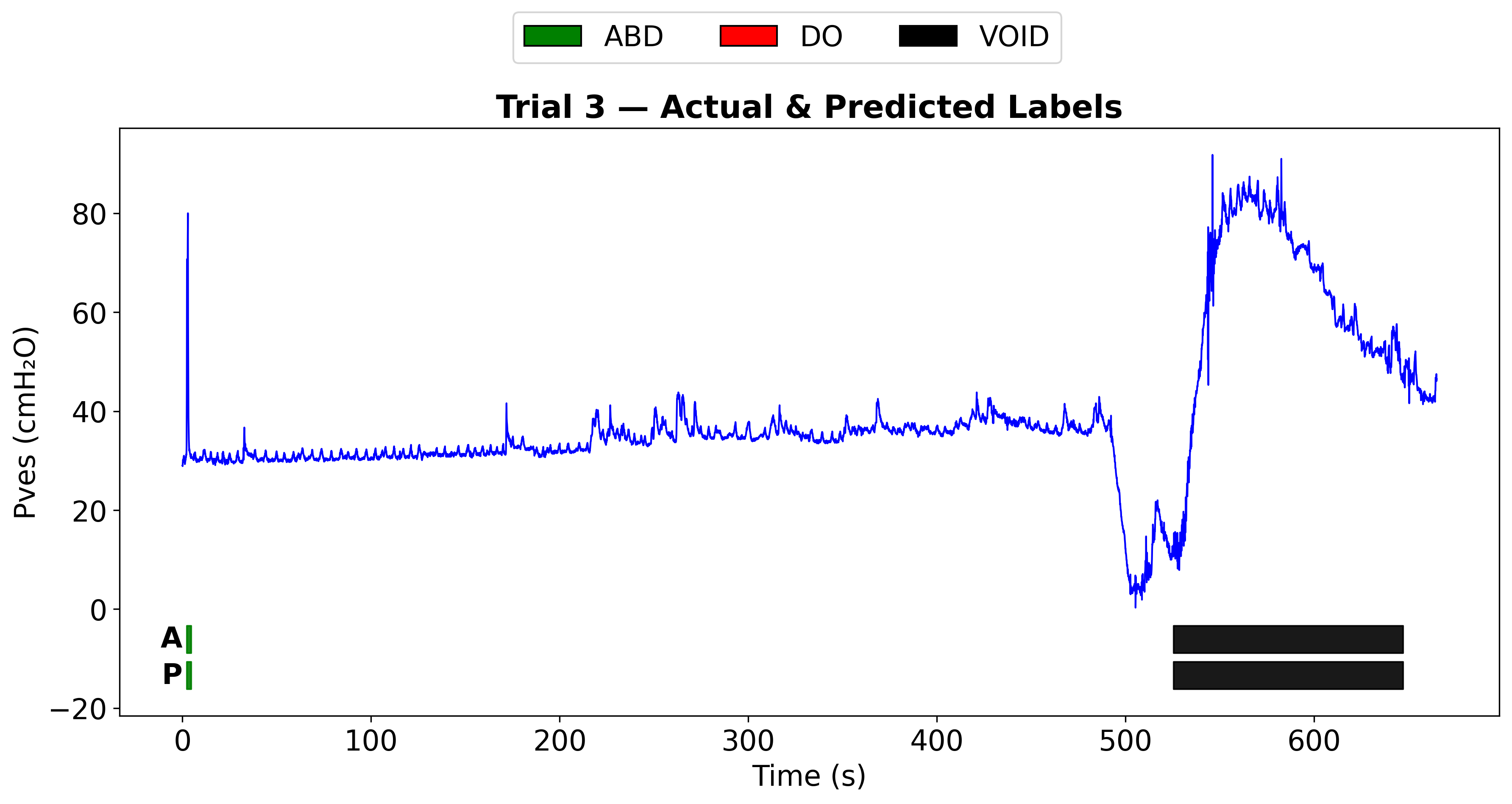}
        \caption{Patient 7}
    \end{subfigure}
    \begin{subfigure}[t]{0.49\textwidth}
        \centering
     \includegraphics[width=\textwidth]{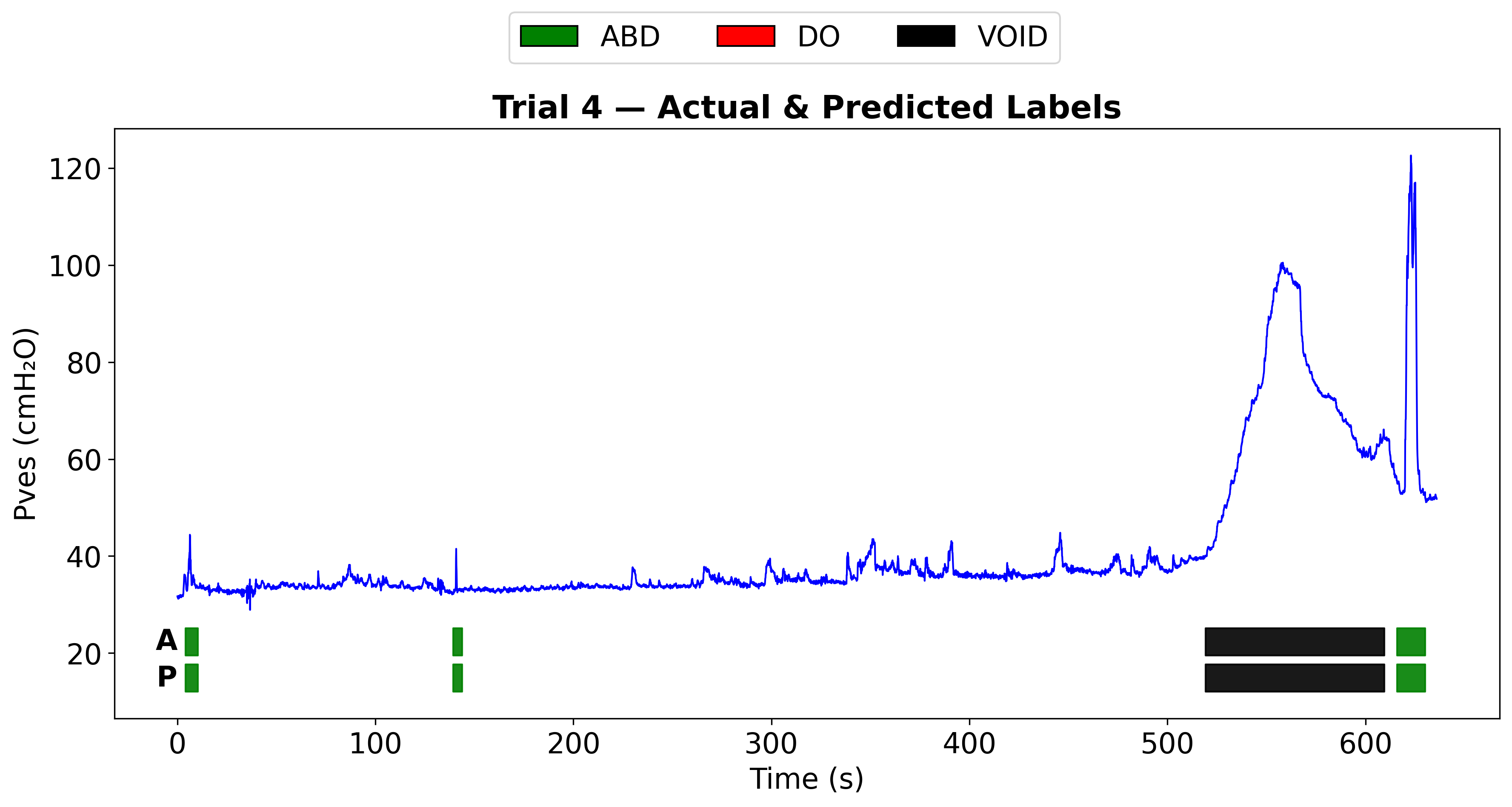}
        \caption{Patient 8}
    \end{subfigure}

        % ---- Row 2 ----
    \begin{subfigure}[t]{0.49\textwidth}
        \centering
       \includegraphics[width=\textwidth]{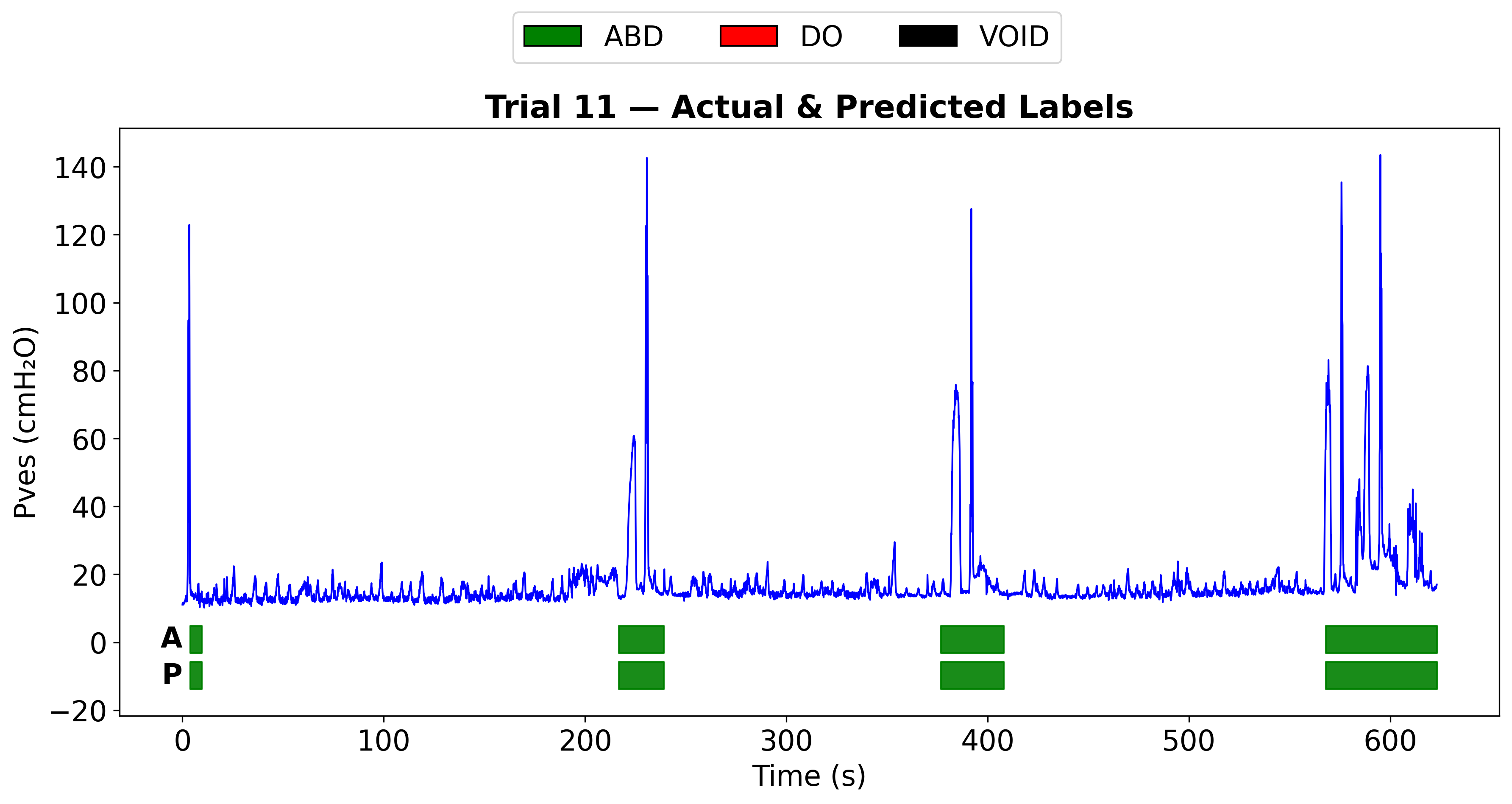}
        \caption{Patient 9}
    \end{subfigure}
    \begin{subfigure}[t]{0.49\textwidth}
        \centering
        \includegraphics[width=\textwidth]{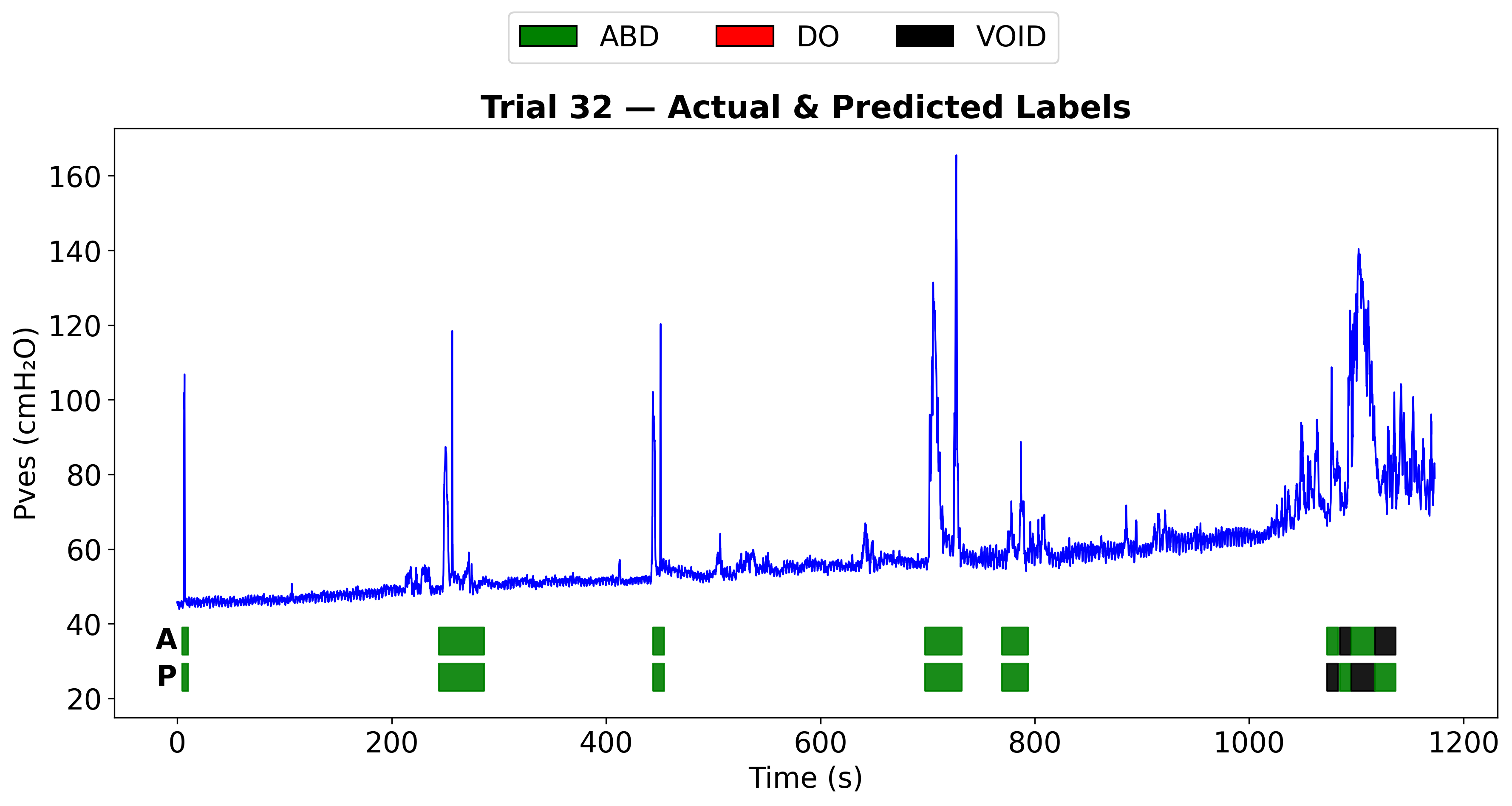}
        \caption{Patient 10}
    \end{subfigure}

    \caption{Actual (A) and predicted (P) event annotations over bladder pressure traces from four subjects in external validation Dataset C, with varying signal morphologies and noise levels. Event types are color-coded as follows: abdominal activity (ABD, green), detrusor overactivity (DO, red), and voiding contractions (VOID, black). The model was trained on Datasets A and B.} 
    \label{fig:pves_actual_pred}
\end{figure*}

\clearpage
\twocolumn